\documentclass[12pt,draftclsnofoot, onecolumn]{IEEEtran}

\IEEEoverridecommandlockouts
\usepackage{cite}
\usepackage{amsmath,amssymb,amsfonts}
\usepackage{algorithmic}
\usepackage{graphicx}
\usepackage{textcomp}
\usepackage{epstopdf}
\usepackage[font=small,skip=0pt]{caption}
\usepackage{multirow}
\usepackage[utf8]{inputenc}
\usepackage[english]{babel}

\newtheorem{theorem}{Theorem}
\newtheorem{lemma}{Lemma}
\newtheorem{definition}{Definition}

\usepackage[usenames, dvipsnames]{color}
\newcommand{\snr}{\mathsf{SNR}}
\newcommand{\sep}{\mathsf{SEP}}

\newcommand{\defeq}{\ensuremath{\triangleq}} 
\newcommand{\dvo}{\mathsf{DVO}}
\newcommand{\re}[1]{#1_{\rm re}}
\newcommand{\im}[1]{#1_{\rm im}}
\newcommand{\BO}[1]{O\paren{#1}} 
\newcommand{\LO}[1]{o\paren{#1}}
\newcommand{\TO}[1]{\Theta\paren{#1}}
\newcommand{\OO}[1]{\Omega\paren{#1}}
\newcommand{\eeq}{\stackrel{\rm e}{=}}
\newcommand{\egeq}{\stackrel{\rm e}{\geq}}
\newcommand{\eleq}{\stackrel{\rm e}{\leq}}
\newcommand{\Arg}[1]{\mathsf{Arg}\left(#1\right)}
\newcommand{\dist}[1]{\mathsf{dist}\left(#1\right)}
\newcommand{\paren}[1]{\left(#1\right)}
\newcommand{\sqparen}[1]{\left[#1\right]}
\newcommand{\brparen}[1]{\left\{#1\right\}}

\newcommand{\parenro}[1]{\left.\left[#1\right)\right.}
\newcommand{\abs}[1]{\left| #1\right|}
\newcommand{\field}[1]{\ensuremath{\mathbb{#1}}}
\newcommand{\N}{\ensuremath{\field{N}}} 
\newcommand{\R}{\ensuremath{\field{R}}} 
\newcommand{\C}{\ensuremath{\field{C}}} 
\newcommand{\ra}{\ensuremath{\rightarrow}} 
\newcommand{\PR}[1]{\ensuremath{\mathsf{Pr}\left\{#1\right\}}} 
\newcommand{\ES}[1]{\ensuremath{\mathsf{E}\left[#1 \right]}} 
\newcommand{\V}[1]{\ensuremath{\mathsf{Var}\left(#1 \right)}} 
\newcommand{\e}[1]{\ensuremath{{\rm e}^{#1}}} 

\def\BibTeX{{\rm B\kern-.05em{\sc i\kern-.025em b}\kern-.08em
    T\kern-.1667em\lower.7ex\hbox{E}\kern-.125emX}}
\begin{document}

\title{Low-Resolution Quantization in Phase Modulated Systems: Optimum Detectors and Error Rate Analysis}

\author{Samiru Gayan, Rajitha Senanayake, Hazer Inaltekin and Jamie Evans
\thanks{Samiru Gayan, Rajitha Senanayake and Jamie Evans are with the Department of Electrical and Electronic Engineering, The University of Melbourne, Parkville, VIC 3100, Australia.  (e-mail: hewas@student.unimelb.edu.au, rajitha.senanayake@unimelb.edu.au, jse@unimelb.edu.au.)

Hazer Inaltekin is with the School of Engineering, Macquarie University, North Ryde, NSW 2109, Australia. (e-mail: hazer.inaltekin@mq.edu.au)
}
}
\maketitle

\begin{abstract}
This paper considers a wireless communication system with low-resolution quantizers, in which transmitted signals are corrupted by fading and additive noise. For such wireless systems, a {\em universal} lower bound on the average symbol error probability ($\sep$), correct for all $M$-ary modulation schemes, is obtained when the number of quantization bits is not enough to resolve $M$ signal points. In the special case of $M$-ary phase shift keying ($M$-PSK), the optimum maximum likelihood detector for equi-probable signal points is derived. Utilizing the structure of the derived optimum receiver, a general average $\sep$ expression for $M$-PSK modulation with $n$-bit quantization is obtained when the wireless channel is subject to fading with a circularly-symmetric distribution. Adopting this result for Nakagami-$m$ fading channels, easy-to-evaluate expressions for the average $\sep$ for $M$-PSK modulation are further derived.  It is shown that a transceiver architecture with $n$-bit quantization is {\em asymptotically} optimum in terms of communication reliability if $n \geq \log_2M +1$. That is, the decay exponent for the average $\sep$ is the same and equal to $m$ with infinite-bit and $n$-bit quantizers for $n\geq \log_2M+1$. On the other hand, it is only equal to $\frac12$ and $0$ for $n = \log_2M$ and $n < \log_2M$, respectively. An extensive simulation study is performed to illustrate the derived results and energy efficiency gains obtained by means of low-resolution quantizers. 
\end{abstract}


\section{Introduction} \label{sec1}
\subsection{Background and Motivation}
Analog-to-digital converters (ADCs) are known to consume most of the power dissipated at a base station \cite{Bai15}. It is shown that the power consumed by ADCs grows exponentially with their resolution level and linearly with their sampling rate \cite{paper_40,paper_46}. Thus, using high-resolution quantization with high sampling rates can significantly degrade the energy efficiency of a communication system. With the introduction of massive multiple-input-multiple-output (MIMO) and millimeter wave (mmWave) technology, this is even more prominent in next generation wireless systems. Because, massive MIMO systems use hundreds of antennas where each antenna is connected to a dedicated radio frequency (RF) chain equipped with high-resolution ADCs. MmWave systems, on the other hand,  use much larger bandwidths that require higher sampling rates.  In fact, the typical power consumption of a high speed ($\geq 20$ GSamples/s) and high-resolution ($8$-$12$ bits) ADC is around $500$ [mWatts]. Therefore, a future mmWave massive MIMO system with $256$ RF chains and $512$ ADCs will require around $256$ [Watts] of power \cite{paper_55}, which is potentially unaffordable. Consequently, the idea of replacing power hungry high-resolution ADCs with low-resolution ADCs could provide a viable solution to the power consumption concerns in future wireless systems.

Indeed, low-resolution ADCs have long been known to provide significant energy savings in digital transceiver implementations \cite{paper_22,paper_23, Madow09}. Their other benefits include simplification in design (especially with $1$-bit ADCs) and reduction in transceiver form-factor \cite{Dabeer10,paper_20,paper_55}. Furthermore, the future long-term evolution (LTE) networks are also expected to support a wide range of Internet-of-Things (IoT) applications through protocols such as LTE-M, NB-IoT and EC-GSM, where devices are usually battery power-limited \cite{IoT5g}. In these future application scenarios, low-resolution ADC based digital transceivers have the potential to prolong the battery lifetime of remote IoT devices as well, and thereby lessening the operating costs and the need for frequent human intervention.

This paper investigates the performance of a wireless communication system with low resolution ADCs, in a symbol error probability $\paren{\sep}$ perspective. In our analysis, we consider the optimum maximum likelihood (ML) detector, and to provide a thorough discussion, we focus on single-input single-output (SISO) channels. Most of the previous work on low-resolution ADCs have focused on the abstract case of $1$-bit quantization \cite{paper_3, paper_8, paper_20, paper_25, paper_22, Lee17, paper_32, paper_34, paper_33, paper_26, paper_29, paper_27, paper_28, paper_30}, where a simple comparator forwards the sign of the signal to the digital domain and discards all the information about the analog signal amplitude.  Such comparators consume negligible power and does not require an automatic gain control circuit. Thus, they lead to cost and power effective implementation of RF chains \cite{Lee17}.
In this paper, we take a different approach in which we allow the number of bits in the quantizer to vary until the transceiver architecture becomes asymptotically optimum in terms of communication reliability. Focusing on a special phase quantizer, we derive analytical expressions of the average $\sep$ when the channels are subject to Nakagami-$m$ fading and the transmitted bits are modulated using $M$-PSK modulation. More importantly, our asymptotic results reveal a fundamental ternary behaviour in the error probability performance of a wireless communication system with low-resolution ADCs, providing an important insight to system designers when choosing the required number of quantization levels.

Using low-resolution ADCs in wireless communication systems has been investigated in various aspects. The performance of communication systems with low-resolution ADCs is lower than that of the idealized systems without quantization or traditional systems with high-resolution ADCs. Therefore, performance analysis of low-resolution quantization is a key research area. It was shown in \cite{paper_26} that the capacity of point-to-point MIMO channel with $1$-bit ADCs is lower bounded by the rank of the channel in the high signal-to-noise ratio ($\snr$) regime. Results in \cite{paper_27} and \cite{paper_28} show that the channel capacity reduces by a factor of $2/\pi$ (1.96 dB) in the low-$\snr$ regime for a MIMO system with $1$-bit ADCs, when compared to a conventional high-resolution system. Further, the results in \cite{paper_30} establish the fact that the performance loss due to employing $1$-bit ADCs can be overcome by having approximately 2.5 times more antennas at the base station. \cite{paper_31} focuses on the information rate of a quantized block non-coherent channel with $1$-bit ADCs. The results in this paper show that around $80-85\%$ of the mutual information attained with unquantized observations can also be attained with $3$-bit quantization for QPSK modulation and $\snr$ greater than $2$-$3$ dB. In \cite{paper_36}, Liang {\em et al.} presented a mixed-ADC architecture for MIMO systems in which some of the high-resolution ADCs were replaced with $1$-bit ADCs. Their results show that the proposed architecture can achieve a near-similar performance as conventional architecture while reducing the energy consumption considerably.

Signal detection rules developed for receivers with high-resolution ADCs often become sub-optimal for receivers with low-resolution ADCs \cite{paper_55}. In \cite{Mezghani_12}, the authors propose a linear minimum mean square error (LMMSE) receiver when in-phase and quadrature components of the received signal are independently quantized by using a low-resolution ADC. They provide an approximation for the mean squared error between the transmitted symbol and the received one, and derive an optimized linear receiver which performs better than the conventional Weiner filter. Results in \cite{Mezghani_12} were further extended to an iterative decision feedback receiver with quantized observations in \cite{paper_13}. For the same quantizer structure of independent quantization of in-phase and quadrature signal components, an ML detector was obtained  in \cite{paper_3} by using {\em only} $1$-bit ADCs. The  complexity of the ML detector proposed in \cite{paper_13} grows exponentially with high signal constellations, number of transmit antennas and network size, which is not practical for real-world deployments. To overcome this difficulty, a near-optimum ML detector was proposed in \cite{paper_8} by using the convex optimization techniques. Although the $\sep$ performance of the proposed near-optimum ML detector is better than the performance of linear detectors, it has been numerically observed that the proposed near-optimum ML detector still suffers from an error floor as $\snr$ increases \cite{paper_55, paper_8}. Complementing this critical observation, in our work, we show the existence of a universal error floor below which the average $\sep$ cannot be pushed down for any $M$-ary modulation scheme and quantizer structure if the number of quantization bits is less than $\log_2M$.

\subsection{Main Contributions} 
In this paper, we consider a  point-to-point wireless communication system, where data transmission is corrupted by fading and noise. 
%
%
Motivated by the capacity achieving property of circularly symmetric input distributions for low-resolution ADCs \cite{Krone10}, we assume that the transmitted symbols are modulated using $M$-ary phase shift keying ($M$-PSK).  For such a system, we design a low-resolution ADC that quantizes the phase of the received signal in such a way that only the information about the quantization region in which the received signal landed is sent to the detector. The use of phase quantization in our model is further motivated by the following two factors. First, considering channel impairments as phase rotations in transmitted signals, quantization and decision regions for $M$-PSK modulation are conveniently modelled as convex cones in the complex plane \cite{book_5}, and without requiring the use of automatic gain control. Second, phase quantizers can be implemented using $1$-bit ADCs that consist of simple comparators, and they consume negligible power (in the order of mWatts). 
%
%
Our main contributions are summarized as follows.
\begin{itemize}
\item For any $M$-ary modulation scheme and quantizer structure, we show the existence of an error floor below which the average symbol error probability ($\sep$) cannot be pushed if the number of quantization bits $n$ is less than $\log_2 M$.   
\item For $M$-PSK modulation with $M \geq 2$, we derive the optimum ML detection rule for signal detection with low-resolution ADCs.
\item We obtain analytical expressions for the average $\sep$ attained by the derived ML rule with $n$-bit quantization when the wireless channel is subjected to Nakagami-$m$ fading. 
\item We establish a fundamental ternary average $\sep$ behaviour with low-resolution ADCs and $M$-PSK modulation under the Nakagami-$m$ fading model. In particular, we show that the decay exponent of the average $\sep$ is the same with that of an infinite-bit quantization, which is equal to $m$, when $n$ is larger than or equal to $\log_2M + 1$. We also show that it is equal to $\frac12$ and $0$ for $n=\log_2M$ and $n<\log_2M$, respectively. 
\item We perform a detailed numerical analysis in the high-$\snr$ regime to corroborate the derived analytical results and to illustrate the energy gains obtained by low-resolution ADCs. 
%
%
\end{itemize}

From a system design point of view, our results show that using one additional bit on top of $\log_2 M$ of them can achieve optimum communication robustness in the high-$\snr$ regime. In particular, for fading environments with a large value of $m$, using an extra quantization bit improves communication reliability significantly. On the other hand, it may be more beneficial to use $\log_2 M$ bits for small values of $m$, without sacrificing from communications robustness too much but doubling system energy efficiency.  


\subsection{Notation}
We use uppercase letters to represent random variables and calligraphic letters to represent sets. We use $\R$, $\R^2$ and $\N$ to denote the real line, $2$-dimensional Euclidean space and natural numbers, respectively. For a pair of integers $i \leq j$, we use $\sqparen{i:j}$ to denote the discrete interval $\brparen{i, i+1, \ldots, j}$. For two functions $f$ and $g$, we will say $f(x) = \BO{g(x)}$ as $x \ra x_0$ if $\abs{f(x)} \leq c \abs{g(x)}$ for some $c > 0$ when $x$ is sufficiently close to $x_0$. Similarly, we will say $f(x) = \OO{g(x)}$ as $x \ra x_0$ if $\abs{f(x)} \geq c \abs{g(x)}$ for some $c > 0$ when $x$ is sufficiently close to $x_0$. We write $f(x) = \TO{g(x)}$ as $x \ra x_0$ if $f(x) = \BO{g(x)}$ and $f(x) = \OO{g(x)}$ as $x \ra x_0$. Finally, we will say $f(x) = \LO{g(x)}$ as $x \ra x_0$ if $\lim_{x \ra x_0} \abs{\frac{f(x)}{g(x)}} = 0$.          

The set of complex numbers $\C$ is $\R^2$ equipped with the usual complex addition and complex multiplication. We write $z = \re{z} + \jmath \im{z}$ to represent a complex number $z \in \C$, where $\jmath = \sqrt{-1}$ is the {imaginary unit} of $\C$, and $\re{z}$ and $\im{z}$ are called, respectively, {\em real} and {\em imaginary} parts of $z$ \cite{Remmert91}. Every $z \in \C$ has also a {\em polar} representation $z = \abs{z}\e{\jmath \theta} = \abs{z}\paren{\cos\paren{\theta} + \jmath \sin\paren{\theta}}$, where $\abs{z} \defeq \sqrt{\re{z}^2 + \im{z}^2}$ is the magnitude of $z$ and $\theta = \Arg{z} \in [-\pi, \pi)$ is called the (principle) argument of $z$.\footnote{The range of $\Arg{z}$ can be taken to be any interval of length $2\pi$. For our purposes, taking its range to be $[-\pi, \pi)$ will help to simplify the notation for some integral expressions.} As is common in the communications and signal processing literature, $\Arg{z}$ will also be called the phase of $z$ (modulo $2\pi$).  For a complex random variable $Z = \re{Z} + \jmath \im{Z}$, we define its mean and variance as $\ES{Z} \defeq \ES{\re{Z}} + \jmath \ES{\im{Z}}$ and $\V{Z} \defeq \ES{\abs{Z - \ES{Z}}^2}$, respectively.  We say that $Z$ is {\em circularly-symmetric} if $Z$ and $\e{\jmath \theta}Z$ induce the same probability distribution over $\C$ for all $\theta \in \R$ \cite{Picinbono94,Koivunen12}.
For $x > 0$, $\log x$ and $\log_2 x$ will denote natural logarithm of $x$ and logarithm of $x$ in base $2$, respectively. 

\section{System Setup} 
\label{sec2}
\subsection{Channel Model and Signal Modulation} 
\label{Subsection: Channel Model}
We consider the classical point-to-point wireless channel model with flat-fading. For this channel, the received discrete-time baseband equivalent signal $Y$ can be expressed by   
\begin{equation}\label{eq1}
Y = \sqrt{\snr}H X + W,
\end{equation}
where $X \in \mathcal{C} \subset \C$ is the transmitted signal, $\mathcal{C}$ is the constellation set of information signals in $\C$, $\snr$ is the ratio of the transmitted signal energy to the additive white Gaussian noise (AWGN) spectral density, $H \in \C$ is the unit power channel gain between the transmitter and the receiver, and $W$ is the circularly-symmetric zero-mean unit-variance AWGN, i.e., $W \sim  \mathcal{CN}(0,1)$. In order to formalize the receiver architecture and the optimum signal detection problem below, we will assume that $\mathcal{C}=\brparen{\e{\jmath \pi \paren{\frac{2k +1}{M}-1}}}_{k=0}^{M-1}$ in the remainder of the paper, which is the classical $M$-ary phase shift keying ($M$-PSK) signal constellation\footnote{This choice of $\mathcal{C}$ ensures that the phase of $X$ always lies in $[-\pi, \pi)$} and for ease of exposition, we only consider the case in which $M$ is an integer power of $2$\footnote{Extensions of our results to the more general case of $M$ being any positive integer is straightforward, albeit with more complicated notation and separate analyses in some special cases.}.

\subsection{Receiver Architecture} \label{Subsection: Receiver}
\begin{figure}[!t]
\center
\includegraphics[width=0.8\textwidth]{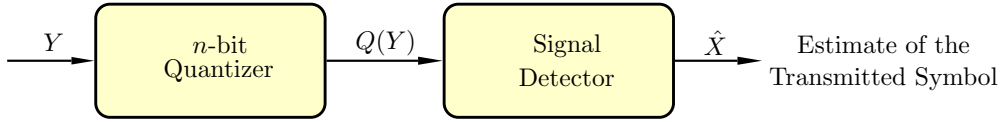}
\hspace{10mm}
\caption{The receiver architecture with low-resolution quantization. The signal detector observes only the $n$-bit quantized versions of $Y$ to estimate the transmitted signal.}
\label{sys_model}
\end{figure}

The receiver architecture is based on a low-resolution ADC. As illustrated in Fig. \ref{sys_model}, the received signal $Y$ is first sent through a low-resolution quantizer, and then the resulting quantized signal information is used to determine the transmitted symbol $X$. More specifically, if $n$ bits are used to quantize $Y$, the quantizer $Q$ divides the complex domain $\C$ into $2^n$ quantization regions and outputs the index of the region in which $Y$ lies as an input to the detector. As such, we declare $Q(Y) = k$ if $Y \in \mathcal{R}_k$ for $k \in \sqparen{0:2^n -1}$, where $\mathcal{R}_k \subseteq \C$ is the $k$th quantization region.  Since information is encoded in the phase of $X$ with the above choice of constellation points, we choose $\mathcal{R}_k$ as the convex cone given by
\begin{eqnarray*}
\mathcal{R}_k = \brparen{z \in \C: \frac{2\pi}{2^n}k \leq \Arg{z} + \pi < \frac{2\pi}{2^n}\paren{k+1}}.
\end{eqnarray*}

We also assume that full channel state information is available at the receiver. The assumption on the availability of full channel state information at the receiver is justified by the previous work on channel estimation with low-resolution ADCs \cite{Dabeer10, Mezghani2010, Mo2018, Wen16}. 
In particular, it was shown in \cite{Dabeer10} that it is possible to attain a near full-precision channel estimation performance with the use of low-resolution ADCs by increasing the number of training symbols in the closed-loop estimation process.  Further, mixed-ADC architectures can also be employed to achieve high channel estimation accuracy \cite{paper_36}.

\section{Optimum Signal Detection} \label{Subsection: Signal Detection}
The aim of the detector is to minimize the $\sep$ by using the knowledge of $Q(Y)$ and channel state information, which can be represented as selecting a signal point $\hat{x}\paren{k, h}$ satisfying 
\begin{eqnarray}
\hat{x}\paren{k, h} \in \underset{x \in \mathcal{C}}{\arg\max }\ \PR{X = x \big| Q(Y) = k, H = h}, \label{Eqn: opt detector}
\end{eqnarray}
for  $h \in \C$ and $k \in \sqparen {0: 2^n-1}$. The main performance figure of merit for the optimum detector is the average $\sep$ given by
\begin{eqnarray}
p\paren{\snr} = \PR{X \neq \hat{x}\paren{Q\paren{Y}, H}}. \label{Eqn: SEP}
\end{eqnarray} 
It is important to note that $p\paren{\snr}$ depends on $\snr$ as well as the number of quantization bits. Our first result indicates that there is an $\snr$-independent error floor such that the average $\sep$ values below which cannot be attained for $n<\log_2M$. The following theorem establishes this result formally.
\begin{theorem} \label{Theorem: lower bound n<log2M}
Let $p_{\min}$ be the probability of the least probable transmitted symbol. If $n <\log_2 M$, then for any choice of modulation scheme and quantizer structure  
\begin{eqnarray}
p\paren{\snr} \geq  \frac{M-2^n}{2^n} p_{\min} \label{Eqn: Universal SEP Lower Bound}
\end{eqnarray} 
for all $\snr \geq 0$. 
\end{theorem}
\begin{IEEEproof}
See Appendix \ref{Appendix:Theorem: lower bound n<log2M}.
\end{IEEEproof}

Firstly, we note that the error floor in \eqref{Eqn: Universal SEP Lower Bound} is always a valid lower bound since $P_{\min} \leq \frac{1}{M}$. Secondly, it does not depend on the fading model. The average $\sep$ values below $\frac{M-2^n}{2^n} p_{\min}$ cannot be achieved due to the inherent inability of low-resolution ADC receivers to resolve different signal points when $n < \log_2M$. We also note that the Fano's inequality can also be used to obtain similar, perhaps tighter, lower bounds on $p\paren{\snr}$ \cite{Cover91}. However, this will require the calculation of equivocation between $X$ and $Q(Y)$ for each choice of modulation scheme and quantizer structure. Hence, it is not clear how the minimization is carried out over the modulation and quantizer selections in this approach.

Next, we will assume that all signal points in $\mathcal{C}$ are equiprobable, with probability $\frac{1}{M}$, and hence the optimum detector in \eqref{Eqn: opt detector} is equivalent to the ML detector given by   
\begin{align}
\hat{x}\paren{k, h} \in \underset{x \in \mathcal{C}}{\arg\max }\ \PR{Q(Y) = k \big| X = x, H = h} \label{Eqn: ML detector Q} 
\end{align}
for  $h \in \C$ and $k \in \sqparen {0: 2^n-1}$. Since  $Y$ is a proper complex Gaussian random variable with mean $\ES{Y} = \sqrt{\snr}hx$ and variance $\V{Y} = 1$, we can write the probability in \eqref{Eqn: ML detector Q} as     
\begin{eqnarray} \label{Eqn: Prob of landing}
\PR{Q(Y) = k \big| X = x, H = h} = \int_{\mathcal{R}_k} \frac{1}{\pi} \exp\paren{-\abs{y - \sqrt{\snr}hx}^2} dy, \label{Eqn: ML Probability}
\end{eqnarray}
 where the integral in \eqref{Eqn: ML Probability} is with respect to the standard Borel measure in $\C$ \cite{Massey93}.  The next theorem describes the operation of the ML detector for the above signal detection problem. 
  
\begin{theorem} \label{Theorem: ML Detector}
Assume $H$ has a continuous probability density function (pdf). Then, $\hat{x}\paren{k, h}$ is unique with probability one, i.e., the set of $h$ values for which $\underset{x \in \mathcal{C}}{\arg\max }\ \PR{Q(Y) = k \big| X = x, H = h}$ is singleton has probability one, and the ML detection rule for the low-resolution ADC based receiver architecture can be given as   
\begin{eqnarray}
\hat{x}\paren{k, h} = \underset{x \in \mathcal{C}}{\arg\min} \ \dist{\sqrt{\snr}hx, \mathcal{H}_k},
\end{eqnarray}
where $h \in \C$, $k \in \sqparen{0: 2^n-1}$, $\dist{z, \mathcal{A}}$ is the distance between a point $z \in \C$ and a set $\mathcal{A} \subseteq \C$, which is defined as $\dist{z, \mathcal{A}} \defeq \inf_{s \in \mathcal{A}} \abs{z - s}$, and $\mathcal{H}_k = \brparen{z \in \C: \Arg{z} + \pi = \frac{\pi}{2^n}\paren{2k+1}}$.   
\end{theorem}

\begin{IEEEproof}
See Appendix \ref{Appendix: Optimum Detector Proof}.
\end{IEEEproof}

We note that the half-hyperplane $\mathcal{H}_k$ in Theorem \ref{Theorem: ML Detector} bisects the $k$th quantization region $\mathcal{R}_k$ into two symmetric regions. Hence, Theorem \ref{Theorem: ML Detector} indicates that the most probability mass is accumulated in the region $\mathcal{R}_k$ when the unit-variance proper complex Gaussian distribution with mean closest to $\mathcal{H}_k$ is integrated over $\mathcal{R}_k$.
%
%
Next we use the structure of the ML detection rule to derive integral expressions for $p\paren{\snr}$ for $M\geq 2$ in Section \ref{Section: SEP}. Further, in order to characterize the communication robustness with low-resolution ADCs in the high $\snr$ regime, we also provide a detailed analysis on the asymptotic decay exponent of $p\paren{\snr}$ in Section \ref{Section: Diversity Order}.   

\section{Average Symbol Error Probability} 
\label{Section: SEP}
\subsection{Symbol Error Probability for $n \geq \log_2 M$}
Let us first obtain a key lemma that simplifies the calculations for deriving $p\paren{\snr}$ when the number of quantization bits is at least $\log_2 M$. Note that this lemma holds for general circularly-symmetric fading processes without assuming any specific functional form.   
\begin{lemma}
\label{Lemma: p(snr) = 2^n P_00}
Let $H = R \e{\jmath \Lambda}$ be a circularly-symmetric fading coefficient with $ R$ and $\Lambda$ denoting the magnitude and the phase of $H$, respectively. Let the joint  pdf of $R$ and $ \Lambda$ be given by $f_{R, \Lambda}\paren{r, \lambda} = \frac{1}{2\pi} f_R\paren{r}$ for $\lambda \in \parenro{-\pi, \pi}$ and $r \geq 0$. Then, $p\paren{\snr}$ is equal to
\begin{align}\label{Eqn: p(SNR) Eqn}
p\paren{\snr} &= \frac{2^{n-1}}{\pi} \int_{\frac{\pi}{M}-\frac{\pi}{2^n}}^{\frac{\pi}{M}+\frac{\pi}{2^n}}\int_{0}^{\infty}\PR{\sqrt{\snr}r \e{\jmath\theta} + W \notin \mathcal{E}}f_{R}\paren{r} \,dr \, d\theta,  
\end{align}
where $\mathcal{E} = \brparen{z \in \C: 0 \leq \Arg{z} < \frac{2\pi}{M}}$. 
\end{lemma}
\begin{IEEEproof}
See Appendix \ref{Appendix:p(snr)=2^nP00 Proof}.
\end{IEEEproof}
 
Using Lemma \ref{Lemma: p(snr) = 2^n P_00}, next we obtain integral expressions for $p\paren{\snr}$ when $H$ is circularly-symmetric with the generalized Nakagami-$m$ fading magnitude. We note that the Nakagami-$m$ fading model characterizes a broad range of fading phenomena ranging from severe to moderate and no fading conditions as $m$ varies over $\parenro{0.5, \infty}$ \cite{Nakagami60, Stuber:2001} and it reduces to Rayleigh fading for $m=1$. 

Considering these advantages, we will focus on the Nakagami-$m$ fading model for $H$ to derive integral expressions for $p\paren{\snr}$ in the remainder of the paper. This will be done so for all parameter combinations of $M \geq 2$ (as an integer power of $2$), $n \geq \log_2 M$ and $m \geq 0.5$. It will be seen that the derived integral expressions are easy to calculate numerically and they reduce to simple closed-form expressions in some special cases. Further, we will also show that using $\log_2 M + 1$ bits is enough to achieve the maximum communication robustness achieved by using infinite number of quantization bits. 

\begin{theorem} \label{Theorem: General SEP}
Assume $H$ is a unit-power fading coefficient distributed according to a circularly-symmetric distribution with Nakagami-$m$ fading magnitude. Let $\mathcal{Q}\paren{\cdot}$ be the complementary distribution function of the standard normal random variable and $\Gamma\paren{\cdot}$ be the gamma function \cite{book_1}. Then, for $n \geq \log_2 M$ and $M \geq 2$, $p\paren{\snr}$ is given according to 
\small
\begin{align}\label{Eqn:General SEP} 
p\paren{\snr} &= \left\{ \begin{array}{ll}
			p_1\paren{\snr} + p_2\paren{\snr} - p_3\paren{\snr} + p_4\paren{\snr} & M\geq 4 \\
            p_2\paren{\snr} & M=2
        \end{array} \right. ,\mbox{\normalsize where} \\
p_1\paren{\snr} &= \frac{2^{n-1}m^m}{\pi^{2}}\int_{0}^{\frac{\pi}{2}}\int_{\frac{\pi}{M} - \frac{\pi}{2^{n}}}^{\frac{\pi}{M} + \frac{\pi}{2^{n}}} \left( \frac{\snr}{\sin^2\beta}\cos^2\theta + m \right)^{-m}  d\theta d\beta, \label{Eqn: p_1(SNR)}\\
p_2\paren{\snr} &= \frac{2^{n-1}m^m}{\pi^{2}}\int_{0}^{\frac{\pi}{2}}\int_{\frac{\pi}{M} - \frac{\pi}{2^{n}}}^{\frac{\pi}{M} + \frac{\pi}{2^{n}}} \left( \frac{\snr}{\sin^2\beta}\sin^2\theta + m \right)^{-m}  d\theta d\beta, \label{Eqn: p_2(SNR)} \\
p_3\paren{\snr} &=\frac{2^{n-1}m^m}{\pi^{3}}\int_{0}^{\frac{\pi}{2}}\int_{0}^{\frac{\pi}{2}}\int_{\frac{\pi}{M} - \frac{\pi}{2^{n}}}^{\frac{\pi}{M} + \frac{\pi}{2^{n}}} \paren{ \frac{\snr \cos^2\theta}{\sin^2\beta} + \frac{\snr\sin^2\theta}{\sin^2\gamma} + m }^{-m} d\theta d\beta d\gamma, \label{Eqn: p_3(SNR)} \\
p_4\paren{\snr} &=\frac{2^{n}m^m}{\pi\sqrt{\pi}\Gamma\paren{m}}\int_{\frac{\pi}{M} - \frac{\pi}{2^{n}}}^{\frac{\pi}{M} + \frac{\pi}{2^{n}}}\int_{0}^{\infty}\int_{-\sqrt{\snr}r \cos \lambda}^{\infty}\mathcal{Q}\paren{\sqrt{2\snr}r\sec\paren{\frac{2\pi}{M}}\sin\paren{\frac{2\pi}{M}-\theta} + \sqrt{2}w\tan\paren{\frac{2\pi}{M}}} \label{Eqn: p_4(SNR)} \nonumber \\
& \hspace{9.5cm}\cdot \exp\paren{-\paren{w^2 + mr^2}}dwdrd\theta.
\end{align}
\normalsize         
\end{theorem}

\begin{IEEEproof}
In the following we provide the proof for $M\geq 4$. Please note that the proof for $M=2$ is similar and simpler.
With a slight abuse of notation, we define 
\begin{align} \label{Eqn: p(SNR, h) Expression 1}
p\paren{\snr, h} = \PR{\sqrt{\snr}r \,\e{\jmath\theta} + W \notin \mathcal{E}}, 
\end{align}
where the set $\mathcal{E}$ is defined as in Lemma \ref{Lemma: p(snr) = 2^n P_00}. The probability in \eqref{Eqn: p(SNR, h) Expression 1} can be calculated by 
conditioning on the real part of $W$, which is denoted by $\re{W}$.
%
%
By using Fig. \ref{SEP calculation} as a visual guide,
we can write $p\paren{\snr,h}$ after conditioning on $\re{W}$ as 
%
\begin{figure}[!t]
\center
\includegraphics[width=0.7\textwidth]{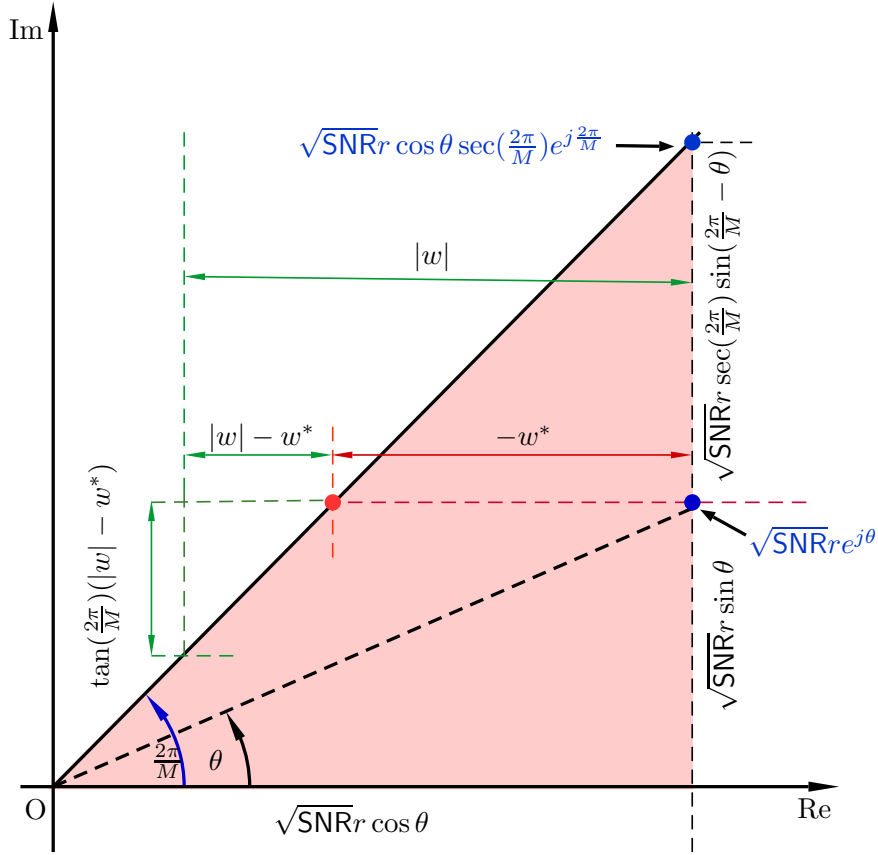}
\caption{An illustration of average $\sep$ calculations. If the noise does not drag the original $M$-PSK constellation point rotated by the channel $h$ beyond the region $\mathcal{E}$ (shaded area), there will not be any errors in decoding.} 
\label{SEP calculation}
\end{figure}
 \begin{align}
\PR{\sqrt{\snr}r\e{j\theta} + W \notin \mathcal{E} \,\big|\, \re{W} = w} \nonumber \\ 
& \hspace{-6cm}= \mathcal{Q}\paren{\sqrt{2\snr} r \sin\theta} +\mathcal{Q}\paren{\sqrt{2\snr}r\sec\paren{\frac{2\pi}{M}}\sin\paren{\frac{2\pi}{M}-\theta} + \sqrt{2}w\tan\paren{\frac{2\pi}{M}}} \label{Eqn: p(SNR, h) Expression 3} 
\end{align}
for $w \geq -\sqrt{\snr} r \cos \theta$. Similarly,  for  $w < - \sqrt{\snr} r \cos\theta$, we get  
\begin{eqnarray}
\PR{\sqrt{\snr}r\e{j\theta} + W \notin \mathcal{E} \,\big|\, \re{W} = w} = 1. \label{Eqn: p(SNR, h) Expression 2}
\end{eqnarray}

Integrating \eqref{Eqn: p(SNR, h) Expression 3} and \eqref{Eqn: p(SNR, h) Expression 2} with respect to the pdf of $\re{W}$, which is given by $f_{\re{W}}\paren{w} = \frac{1}{\sqrt{\pi}}\e{-w^2}$,
%
%
we obtain $p\paren{\snr,h}$ as
\begin{align}\label{Eqn: Conditional SEP 2}
p\paren{\snr,h} \nonumber \\
&\hspace{-1.5cm}= \mathcal{Q}\paren{\sqrt{2\snr}r\cos \theta} + \mathcal{Q}\paren{\sqrt{2\snr}r\sin \theta}-\mathcal{Q}\paren{\sqrt{2\snr}r\cos \theta}\mathcal{Q}\paren{\sqrt{2\snr}r\sin \theta} \nonumber \\
&\hspace{-1.5cm}+ \frac{1}{\sqrt{\pi}}\int_{-\sqrt{\snr}r\cos \theta}^{\infty}\mathcal{Q}\paren{\sqrt{2\snr}r\sec\paren{\frac{2\pi}{M}}\sin\paren{\frac{2\pi}{M}-\theta} + \sqrt{2}w\tan\paren{\frac{2\pi}{M}}} \,\e{-w^2}\,dw.
\end{align}
For Nakagami-$m$ fading distribution with shape parameter $m \geq 0.5$ and spread parameter $\Omega > 0$ \cite{paper_49}, we can write the pdf of the fading magnitude as $f_{R}\paren{r} = \frac{2m^m}{\Gamma(m)\,\Omega^m}r^{2m-1}\e{-\frac{m}{\Omega}r^2}$ for $r \geq 0$. We set $\Omega = 1$ in our calculations to make sure that $H$ has unit-power. We average $p\paren{\snr,h}$ over the fading distribution and solve the resulting integral based on Lemma \ref{Lemma: p(snr) = 2^n P_00}, and the fact that $\theta$ lies between $0$ and $\frac{2\pi}{M}$, to obtain $p\paren{\snr}$ in Theorem \ref{Theorem: General SEP}. 
\end{IEEEproof}

\subsection{Centering Property: Impact of Quantization Bits on the Average $\sep$ }
In this subsection, we will present an intuitive explanation as to why $p\paren{\snr}$ improves with increasing number of quantization bits. In particular, we will observe that one extra bit, on top of $\log_2 M$ of them, provides a desirable centering property that steers the received signal away from the error-prone decision boundaries. This intuition will help to understand the underlying dynamics leading to the ternary behaviour for the decay exponent of $p\paren{\snr}$ that we establish in the high $\snr$ regime in Section \ref{Section: Diversity Order}.   

For $i\in \sqparen{0:M-1}$, let $x_i = \e{\jmath \pi\paren{\frac{2i+1}{M}-1}}$ be the $i$th signal point in the constellation set $\mathcal{C}$ and $\mathcal{E}_i = \brparen{z \in \C: \Arg{x_i} - \frac{\pi}{M} \leq \Arg{z} < \Arg{x_i} + \frac{\pi}{M}}$. It can be shown (i.e., see Appendix \ref{Appendix:p(snr)=2^nP00 Proof}) that the regions defined by $\mathcal{E}_{i,k} \defeq \exp\paren{\jmath\paren{k-2^{n-1}}\frac{2\pi}{2^n}} \mathcal{E}_i$ for $i \in \sqparen{0:M-1}$ and $k \in \sqparen{0:2^n-1}$ contains all $\mathcal{H}_k$'s to which $\sqrt{\snr}hx_i$ is the closest for $h \in \mathcal{D}_k$, where 
\begin{eqnarray*}
\mathcal{D}_0 = \brparen{z \in \C: \pi - \frac{\pi}{2^n} \leq \Arg{z} < \pi} \bigcup \brparen{z \in \C: -\pi \leq \Arg{z} < \frac{\pi}{2^n} - \pi}
\end{eqnarray*}
and 
\begin{eqnarray*}
\mathcal{D}_k = \brparen{z \in \C: \paren{2k -1}\frac{\pi}{2^n} \leq \Arg{z} +\pi < \paren{2k+1}\frac{\pi}{2^n}}.
\end{eqnarray*}
This means that all the received signal points in $\mathcal{E}_{i,k}$ will be detected as $x_i$, and hence $\mathcal{E}_{i,k}$ can be considered as the {\em region of attraction} of $x_i$. This also means that if the received signal lands in $\mathcal{E}_{i,k}$ when $x_i$ is transmitted, then there will not be any detection errors.

\begin{figure}[!t]
\center
\includegraphics[width=0.7\textwidth]{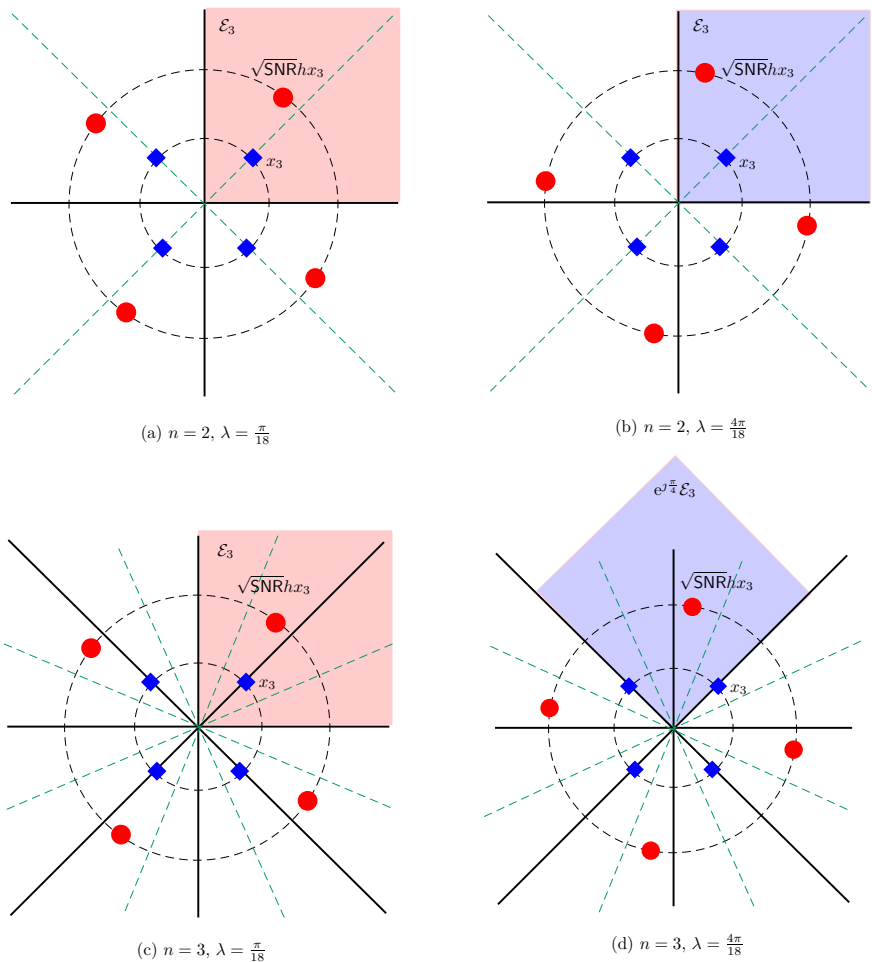}
\caption{An illustration of the centering property for QPSK modulation with $2$-bit and $3$-bit quantization. Original signal points are indicated by `$\diamond$', whereas the rotated ones after multiplication with $\sqrt{\snr}$ and $h$ are indicated by `$\bullet$'. Quantization region boundaries and the corresponding bisectors are indicated in solid black lines and green dash lines, respectively. The shaded area represents the region of attraction of the transmitted symbol $x_3$.}
\label{SEP calculation QPSK}
\end{figure}

Let us consider an example for QPSK modulation with $2$-bit and $3$-bit quantization. Without loss of generality, we will assume that $x_3 = \e{\jmath \frac{\pi}{M}}$ is the transmitted signal. Our analysis will be for two cases of $\lambda = \frac{\pi}{18}$ and $\lambda = \frac{4\pi}{18}$, where $\lambda = \Arg{h}$. Table \ref{Table: Properties} summarizes these two cases, and Fig. \ref{SEP calculation QPSK} illustrates them. In this figure, we show both the original signal points (indicated by `$\diamond$') and the rotated ones (indicated by `$\bullet$') after multiplying with $\sqrt{\snr}$ and $h$.    
\begin{table}
\begin{center}
\begin{tabular}{ |c|c|c|c| } 
\hline
& $\lambda = \frac{\pi}{18}$ & $\lambda = \frac{4\pi}{18}$ \\
\hline
\multirow{3}{4em}{$n=2$} & $h \in \mathcal{D}_2$ & $h \in \mathcal{D}_4$ \\ 
& $i=3$,  $k=2$ & $i=3$,  $k=4$ \\ 
& $\mathcal{E}_{i,k} = \mathcal{E}_3$ & $\mathcal{E}_{i,k} = \mathcal{E}_3$ \\ 
\hline
\multirow{3}{4em}{$n=3$} & $h \in \mathcal{D}_2$ & $h \in \mathcal{D}_4$ \\ 
& $i=3$, $k=2$ & $i=3$, $k=5$ \\ 
& $\mathcal{E}_{i,k} = \mathcal{E}_3$ & $\mathcal{E}_{i,k} = \e{\jmath\frac{\pi}{4}}\mathcal{E}_3$ \\ 
\hline
\end{tabular}
\end{center}
\caption{Centering property for QPSK modulation with $2$-bit and $3$-bit quantization. $\mathcal{E}_{i,k}$ is the region of attraction of the symbol $x_i$ when the quantizer output $Q\paren{Y}=k$.} \label{Table: Properties}
\end{table}

For both $2$-bit and $3$-bit quantization, we observe that $h \in \mathcal{D}_2$ and $h \in \mathcal{D}_4$ for $\lambda = \frac{\pi}{18}$ and $\lambda = \frac{4\pi}{18}$, respectively. Therefore, for $2$-bit quantization, the region of attraction for $x_3$ will be $\mathcal{E}_3$ for both cases. Here, we can see that the rotated constellation point $\sqrt{\snr}hx_3$ is very close to the decision boundary when $\lambda = \frac{4\pi}{18}$. Hence, there is a high probability that the received signal $\sqrt{\snr}hx_3 + w$ lands in the adjacent quantization region for $\lambda = \frac{4\pi}{18}$. In this instance, we will have a detection error. However, with the addition of one bit to the quantizer (i.e. with $3$-bit quantization), the region of attraction of $x_3$ will be $\e{\jmath\frac{\pi}{4}}\mathcal{E}_3$, and hence the ML detector can correctly decode the transmitted signal even if the received one lands in the adjacent quantization region. This is illustrated in Fig. \ref{SEP calculation QPSK}(d).  Therefore, the addition of one extra bit to the quantizer, steers the received signal away from the error-prone decision boundaries to improve $p\paren{\snr}$. Similarly, when the number of bits in the quantizer continues to increase, the quantization regions will become thinner,  and hence the regions of attraction will be better centered around the received signal points. This is the fundamental phenomenon that explains why the average $\sep$ improves with a larger number of quantization bits.

\section{The Decay Exponent for the Average Symbol Error Probability} \label{Section: Diversity Order}
In this section, we will analyze the communication robustness that can be achieved with low-resolution ADCs by focusing on the decay exponent for $p\paren{\snr}$, which is given by\footnote{We will show that the limit in \eqref{Eqn: DVO Definition} exists, and hence there is no ambiguity in the definition of $\dvo$.}
\begin{eqnarray}\label{Eqn: DVO Definition}
\dvo = - \lim_{\snr \ra \infty} \frac{\log{p\paren{\snr}}}{\log{\snr}}. 
\end{eqnarray}
Following the convention in the field, we will call $\dvo$ the {\em diversity order}, although there is only a single diversity branch in our system. It should be noted that Nakagami-$m$ amplitude distribution can be obtained as the envelope distribution of $m$ independent Rayleigh faded signals for integer values of $m$ \cite{Nakagami60}. Hence, visualizing a Nakagami-$m$ wireless channel as a pre-detection analog square-law diversity combiner will put the results of this section into context. We devote the rest of the current section to the proof of this important finding. We will first start with a definition that will simplify the notation below.

\begin{definition} \label{Def: Exponential Equality}
We say a function $f$ is {\em exponentially equal} to $\snr^d$ if $\lim_{\snr \ra \infty} \frac{\log f\paren{\snr}}{\log \snr} = d$ for some $d \in \R$.  We write $f\paren{\snr} \eeq \snr^d$ to indicate exponential equality whenever this limit exists. Similarly, we also write $f\paren{\snr} \eleq \snr^d$ and $f\paren{\snr} \egeq \snr^d$ if $\lim_{\snr \ra \infty} \frac{\log f\paren{\snr}}{\log \snr} \leq d$ and $\lim_{\snr \ra \infty} \frac{\log f\paren{\snr}}{\log \snr} \geq d$, respectively.    
\end{definition}

The following lemma establishes two important properties for exponential equality. 
\begin{lemma} \label{Lemma: sum of pi(SNR) limit}
Let $f\paren{\snr} \eeq \snr^d$ and $f_i\paren{\snr} \eeq \snr^{d_i}$ for $i \in \sqparen{1:N}$. Then, 
\begin{itemize}
\item[(i)] For any $\alpha > 0$, $\alpha f\paren{\snr} \eeq \snr^{d}$ (i.e., invariance with scaling property). 
\item[(ii)] $\sum_{i=1}^N f_i\paren{\snr} \eeq \snr^{d_{\max}}$, where $d_{\max} = \max_{i \in \sqparen{1:N}} d_i$ (i.e., summation property). 
\end{itemize}
\end{lemma}
\begin{IEEEproof}
See Appendix \ref{Appendix: Proof of Lemma: sum of pi(SNR) limit}.
\end{IEEEproof}

The next two lemmas establish the decay rates for $p_1\paren{\snr}$ and $p_2\paren{\snr}$ in Theorem \ref{Theorem: General SEP} in terms of exponential equalities.
\begin{lemma} \label{Lemma: p1(SNR) limit}
For $M \geq 4$, $p_1\paren{\snr}$ is exponentially equal to 
\begin{equation}
p_1\paren{\snr} \eeq
        \left \{ \begin{array}{ll}
            \snr^{-\frac{1}{2}} & \mbox{ if } M=4 \mbox{ and } n=2, \\
            \snr^{-m} & \mbox{ if } M=4 \mbox{ and } n >2, \\
            \snr^{-m} & \mbox{ if } M > 4 \mbox{ and } n \geq \log_2{M}. 
        \end{array} \right .
\end{equation}
\end{lemma}
\begin{IEEEproof}
See Appendix \ref{Appendix:Proof of Lemma: p1(SNR) limit}.
\end{IEEEproof}
\begin{lemma} \label{Lemma: p2(SNR) limit}
For $M \geq 4$, $p_2\paren{\snr}$ is exponentially equal to 
\begin{equation}
p_2\paren{\snr} \eeq
        \left \{ \begin{array}{ll}
            \snr^{-\frac{1}{2}} & \mbox{ if } n = \log_2{M}, \\
            \snr^{-m} & \mbox{ if } n > \log_2{M}. 
        \end{array} \right .
\end{equation}
\end{lemma}
\begin{IEEEproof}
See Appendix \ref{Appendix:Proof of Lemma: p2(SNR) limit}.
\end{IEEEproof}

The following lemma establishes lower and upper bounds on $\sep$ in \eqref{Eqn:General SEP}. We note that the bounds in Lemma \ref{Lemma: SEP bounds} hold for all circularly-symmetric fading processes, including Nakagami-$m$ magnitude pdf as a special case.   
\begin{lemma} \label{Lemma: SEP bounds}
For $M \geq 4$ and $n\geq \log_2{M}$, let $L\paren{\snr} = p_1\paren{\snr}+ \frac{1}{2} p_2\paren{\snr}$ and $U\paren{\snr} = p_1\paren{\snr}+2p_2\paren{\snr}$. Then,   
\begin{align}
L\paren{\snr} \leq p\paren{\snr} \leq U\paren{\snr}. \label{Eqn: Both bounds}
\end{align}
\end{lemma}
\begin{IEEEproof}
See Appendix \ref{Appendix: Proof of Lemma: SEP bounds}.
\end{IEEEproof}
\begin{theorem}\label{Theorem 2}
The $\dvo$ of a low-resolution ADC based receiver architecture with $M$-PSK modulation and Nakagami-$m$ fading is given by \begin{align}\label{diversity_order}
        \dvo =
        \left\{ \begin{array}{ll}
            \frac{1}{2} & n = \log_2 M, \\
            m & n \geq \log_2 M + 1.
        \end{array} \right.
\end{align}
\end{theorem}
\begin{IEEEproof}
The proof for $M \geq 4$ directly follows from Lemmas \ref{Lemma: sum of pi(SNR) limit}, \ref{Lemma: p1(SNR) limit}, \ref{Lemma: p2(SNR) limit} and \ref{Lemma: SEP bounds}. For BPSK modulation (i.e., $M=2$) and $n=1$, we have
\begin{align}
p\paren{\snr} &= \frac{2^{n-1}m^m}{\pi^{2}}\int_{0}^{\frac{\pi}{2}}\int_{0}^{\pi} \paren{\frac{\snr}{\sin^2\beta}\sin^2\theta + m }^{-m}  d\theta d\beta \nonumber \\
&= \frac{2^{n-1}m^m}{\pi^{2}}\int_{0}^{\frac{\pi}{2}}\int_{0}^{\frac{\pi}{2}} \paren{\frac{\snr}{\sin^2\beta}\sin^2\theta + m }^{-m}  d\theta d\beta \nonumber \\
&\hspace{4cm}+ \frac{2^{n-1}m^m}{\pi^{2}}\int_{0}^{\frac{\pi}{2}}\int_{\frac{\pi}{2}}^{\pi} \paren{\frac{\snr}{\sin^2\beta}\sin^2\theta + m }^{-m}  d\theta d\beta  \label{Eqn:p(SNR)_int2}
\end{align}
By using the change of variables $\hat{\theta}=\theta-\frac{\pi}{2}$ in the second integral term of \eqref{Eqn:p(SNR)_int2}, we have
\begin{align}
p\paren{\snr} &= \frac{2^{n-1}m^m}{\pi^{2}}\int_{0}^{\frac{\pi}{2}}\int_{0}^{\frac{\pi}{2}} \paren{\frac{\snr}{\sin^2\beta}\sin^2\theta + m }^{-m}  d\theta d\beta \nonumber \\
&\hspace{4cm}+ \frac{2^{n-1}m^m}{\pi^{2}}\int_{0}^{\frac{\pi}{2}}\int_{0}^{\frac{\pi}{2}} \paren{\frac{\snr}{\sin^2\beta}\cos^2\hat{\theta} + m }^{-m}  d\hat{\theta} d\beta  \label{Eqn: BPSK 1 bit}
\end{align}
This expression is equivalent to $p_1\paren{\snr} + p_2\paren{\snr}$ for $M=4$ and $n=2$. Hence, by using Lemma \ref{Lemma: sum of pi(SNR) limit}, we can conclude that
\begin{align}
\lim\limits_{\snr \ra \infty} - \frac{\log\paren{p\paren{\snr}}}{\log\paren{\snr}}=\frac{1}{2}
\end{align}
for BPSK modulation with $1$-bit quantization and $m\geq \frac{1}{2}$.

For BPSK modulation with $n>\log_2\paren{M}$, we have
\begin{align}
p\paren{\snr} &= \frac{2^{n-1}m^m}{\pi^{2}}\paren{\snr}^{-m}\int_{0}^{\frac{\pi}{2}}\int_{\frac{\pi}{M} - \frac{\pi}{2^{n}}}^{\frac{\pi}{M} + \frac{\pi}{2^{n}}} \paren{\frac{\sin^2\theta}{\sin^2\beta} + \frac{m}{\snr} }^{-m}  d\theta d\beta \label{Eqn: BPSK n>=2}
\end{align}
Therefore
\begin{align}
\log\paren{p\paren{\snr}} &= c - m\log\paren{\snr} + \log\paren{\int_{0}^{\frac{\pi}{2}}\int_{\frac{\pi}{M}-\frac{\pi}{2^n}}^{\frac{\pi}{M}+\frac{\pi}{2^n}} \paren{\frac{\sin^2\theta}{\sin^2\beta} + \frac{m}{\snr}}^{-m} d\theta d\beta}, \nonumber
\end{align} 
where $c = \log\paren{\frac{2^{n-1} m^m}{\pi^2}}$.
Define the function $g_{\snr}\paren{\theta, \beta} \triangleq \paren{\frac{\sin^2\theta}{\sin^2\beta} + \frac{m}{\snr}}^{-m}$, indexed by $\snr$. Since it is positive and increases to the limiting function $g_{\infty}\paren{\theta, \beta}= \paren{\frac{\sin^2\theta}{\sin^2\beta}}^{-m}$ as $\snr$ increases, we can use the monotone convergence theorem \cite{book_7} to write 
\begin{align}
 \lim_{\snr \ra \infty}\log\paren{\int_{0}^{\frac{\pi}{2}}\int_{\frac{\pi}{M}-\frac{\pi}{2^n}}^{\frac{\pi}{M}+\frac{\pi}{2^n}} \paren{\frac{\sin^2\theta}{\sin^2\beta} + \frac{m}{\snr}}^{-m} d\theta d\beta} &= \log\paren{\int_{0}^{\frac{\pi}{2}}\int_{\frac{\pi}{M}-\frac{\pi}{2^n}}^{\frac{\pi}{M}+\frac{\pi}{2^n}} \paren{\frac{\sin^2\theta}{\sin^2\beta}}^{-m} d\theta d\beta}. \nonumber
\end{align}
We note that the last integral is finite since $g_{\infty}\paren{\theta, \beta}$ is continuous and finite over the range of integration. Therefore, for BPSK  modulation with $n>\log_2\paren{M}$, we get
\begin{align}
    \lim\limits_{\snr \ra \infty} - \frac{\log\paren{p\paren{\snr}}}{\log\paren{\snr}} = m.
\end{align}
\end{IEEEproof}

The $\dvo$ analysis above helps to discover the first-order effects of the low-resolution ADC based receivers on the $\sep$ system performance.  In particular, we observe that it is enough to use $\log_2 M + 1$ bits for quantizing the received signal to extract full diversity, which is equal to $m$ for Nakagami-$m$ faded wireless channels. Considering the fact that energy consumption increases exponentially with the number of quantization bits \cite{paper_12}, this finding indicates that a  significant energy saving is possible by means of low-resolution ADC based receivers without any (first order) loss in communication robustness.  

We also observed that the $\dvo$ is only equal to $\frac12$ when $n = \log_2 M$. Together with the universal bound obtained in Theorem \ref{Theorem: lower bound n<log2M}, the discovered {\em ternary} behaviour has significant implications in terms of how to choose the number of quantization bits for low-resolution ADC based receivers. In particular, for fading environments with $m$ close to $\frac12$, a system designer may decide to trade off reliability for energy consumption, without having too much degradation in average $\sep$ by using $\log_2 M$ bits. On the other hand, for fading environments with large $m$, it is more beneficial to use one extra bit to have a major improvement in average $\sep$. 

\section{Performance Analysis for QPSK modulation}
\label{Sec: Effect}
In this section, we conduct a performance analysis for QPSK modulation with low-resolution ADCs by using our results in previous sections. 
%
%
We first present a simplified version of the average $\sep$ expression in \eqref{Eqn:General SEP} for QPSK modulation, and then we analyze the effect of low-resolution quantization under Rayleigh fading.   
\subsection{Symbol Error Probability for QPSK modulation}
{\em Nakagami-$m$ Fading:} In the special case of QPSK modulation (i.e., $M=4$), the average $\sep$ expression in \eqref{Eqn:General SEP} can be further simplified to produce 
\begin{align}\label{Eqn: SEP for QPSK}
p\paren{\snr} = p_1\paren{\snr} + p_2\paren{\snr} - p_3\paren{\snr},
\end{align}
because $\tan\paren{\frac{2\pi}{M}} = \infty$ for $M=4$. By using hypergeometric function ${}_2F_1[\cdot]$ \cite{book_1}, we can simplify \eqref{Eqn: SEP for QPSK} for 2-bit quantization (i.e., $M=4$ and $n=2$) as 
\begin{align}
p\paren{\snr} & = \frac{2}{\pi}\int_{0}^{\frac{\pi}{2}}{}_2F_1 \sqparen{\frac{1}{2},m,1,\frac{-\snr}{m \sin^2\beta}}\,d\beta \nonumber \\
& \hspace{1cm}- \frac{m^m}{\pi}\int_{0}^{\frac{\pi}{2}}\int_{0}^{\frac{\pi}{2}}\paren{\frac{\snr}{\sin^2\gamma}+m}^{-m}{}_2F_1 \sqparen{\frac{1}{2},m,1,\frac{\snr\paren{\sin^2 \beta - \sin ^2 \gamma}}{\snr + m \sin^2 \gamma}}\,d\beta\, d\gamma. \nonumber
\end{align}

{\em Rayleigh Fading:} For special case of Rayleigh fading, which is obtained by setting $m=1$, the expression in \eqref{Eqn: SEP for QPSK} can be re-expressed as 
\begin{align}\label{Eqn: SEP QPSK Rayleigh}
  p\paren{\snr} &= \frac{2^{n}}{\pi^{2}}\int_{0}^{\frac{\pi}{2}}\frac{\sin\beta}{\sqrt{\snr + \sin^{2}\beta}}\arctan\paren{\frac{2\sin\beta\sqrt{\snr + \sin^{2}\beta}}{\snr + 2\sin^{2}\beta}\tan\paren{\frac{\pi}{2^{n-1}}}}d\beta \quad \nonumber \\
   & \hspace{1cm}-\frac{2^{n-1}}{\pi^{3}}\int_{0}^{\frac{\pi}{2}}\int_{0}^{\frac{\pi}{2}} \sqrt{\frac{\sin^{2}\beta \sin^{2}\gamma}{\paren{\snr + \sin^{2}\beta}\paren{\snr + \sin^{2}\gamma}}} \,\cdot \arctan\paren{\vartheta}\,d\beta \,d\gamma,
\end{align}
where
$\vartheta  = \frac{2\sin\beta \sin\gamma \sqrt{(\snr + \sin^{2}\beta)(\snr + \sin^{2}\gamma)}}{\snr(\sin^{2}\beta + \sin^{2}\gamma ) + 2\sin^{2}\beta\sin^{2}\gamma}\tan(\frac{\pi}{2^{n-1}}).\notag$
Furthermore, for $2$-bit quantization with Rayleigh fading (i.e., $M=4$, $n=2$ and $m=1$), we can obtain $p\paren{\snr}$ in closed form as   
\begin{align}
 p\paren{\snr} & = \frac{2}{\pi}\arctan\paren{\frac{1}{\sqrt{\snr}}} - \paren{\frac{1}{\pi}\arctan\paren{\frac{1}{\sqrt{\snr}}}}^{2}. \nonumber
\end{align}
This closed-form analytical expression is very easy to compute without resorting to any numerical integration.   

\subsection{Analysis of Quantization Penalty for QPSK Modulation}
By using the Taylor series expansion for high $\snr$ values, we can re-express the average $\sep$ expressions for QPSK modulation under Rayleigh fading given in \eqref{Eqn: SEP QPSK Rayleigh} as
\begin{equation}\label{Eqn: Asymptotic SEP QPSK Rayleigh}
        p_A\paren{\snr,n} =
        \left\{ \begin{array}{ll}
            \frac{2}{\pi}\snr^{-\frac{1}{2}} + \LO{\snr^{-\frac{1}{2}}} & n = 2 \\
            \frac{2^{n-1}\paren{4\pi - 1}}{\pi^{3}}\tan\paren{\frac{\pi}{2^{n-1}}}\snr^{-1} + \LO{\snr^{-1}} & n \geq 3.
        \end{array} \right.
\end{equation}
While phase quantization  with less number of quantization bits is desirable, due to less processing complexity at the receiver, it deteriorates the average $\sep$ performance of the system. In the following, we quantify the increase in the average $\sep$ as a quantization penalty defined as
\begin{align}\label{Eqn: Quantization Penalty 1}
  \Psi\paren{\snr,n} & = 10\log\paren{\frac{p_A\paren{\snr,n}}{p_A\paren{\snr,\infty}}}, 
\end{align}
where $p_A\paren{\snr,\infty}$ is the average $\sep$ with infinite number of quantization bits. Based on \eqref{Eqn: Asymptotic SEP QPSK Rayleigh}, we can derive $p_A\paren{\snr,\infty}$ as
\begin{align}\label{Eqn: Asymptotic SEP QPSK Rayleigh infinite bits}
  p_A\paren{\snr,\infty} & = \paren{\frac{4\pi -1}{\pi^{2}}}\snr^{-1} + \LO{\snr^{-1}},
\end{align}
where we have used the small-angle approximation $\tan(x) = x$ as $x \rightarrow 0$. 
Substituting \eqref{Eqn: Asymptotic SEP QPSK Rayleigh} and \eqref{Eqn: Asymptotic SEP QPSK Rayleigh infinite bits} into \eqref{Eqn: Quantization Penalty 1} and doing some mathematical manipulations, we can derive the  quantization penalty in terms of average $\sep$ with $n$-bit quantization as
\begin{equation}\label{Eqn: Quantization Penalty 1(a)}
        \Psi\paren{\snr,n} =
        \left\{ \begin{array}{ll}
            10\log \paren{\paren{\frac{2\pi}{4\pi - 1}}\snr^{\frac{1}{2}}} + \LO{\snr^{\frac{1}{2}}} & n = 2 \\
            10\log \paren{\frac{2^{n-1}}{\pi}\tan\paren{\frac{\pi}{2^{n-1}}}} + \LO{1} & n \geq 3.
        \end{array} \right.
\end{equation}
In Section \ref{Sec: Numerical Examples}, we use $\Psi\paren{\snr,n}$ to quantify the increase in average $\sep$ as we change from infinite-bit to $n$-bit quantization.

We further notice that, in order to achieve the same average $\sep$ as with $n$-bit quantization, we need to transmit the signal using a higher power if we use only $(n-1)$-bit quantization. In the following, we quantify the increase in the transmit power as another quantization penalty defined by
\begin{align}\label{Eqn: Quantization Penalty 2}
  \Phi\paren{\sep,n} = 10\log \paren{\frac{\snr_{n-1}}{\snr_{n}}},
\end{align}
where $\snr_{n}$ and $\snr_{n-1}$ are the $\snr$ values required to achieve a certain average $\sep$ with $n$ and $n-1$ quantization bits, respectively. Substituting \eqref{Eqn: Asymptotic SEP QPSK Rayleigh}  into \eqref{Eqn: Quantization Penalty 2} and doing some mathematical manipulations, we can derive the  quantization penalty with $n$-bit  quantization as
\begin{equation}\label{Eqn: Quantization Penalty 2(a)}
\Phi\paren{\sep,n} =
        \left\{\begin{array}{ll}
          10\log \paren{\frac{\pi^{2}}{2(4\pi - 1)}\paren{\snr_{2}}^{\frac{1}{2}}} & n=3\\
          10\log\paren{\frac{1}{2}\tan\paren{\frac{\pi}{2^{n-2}}}\cot\paren{\frac{\pi}{2^{n-1}}}} & n\geq4.
        \end{array}\right.
\end{equation}
In Section \ref{Sec: Numerical Examples}, we use $\Phi\paren{\sep,n}$ to quantify the required transmit power increase as we change from $n$-bit to $(n-1)$-bit quantization.

\section{Numerical Results}\label{Sec: Numerical Examples}
In this section, we present analytical and simulated $\sep$ results for $M$-PSK modulation with $n$-bit quantization.  Channel fading is unit-power and circularly-symmetric with Nakagami-$m$ distributed magnitude, and additive noise is complex Gaussian with zero mean and unit variance. 

\begin{figure}[!t]
\center
\includegraphics[width=0.7\textwidth]{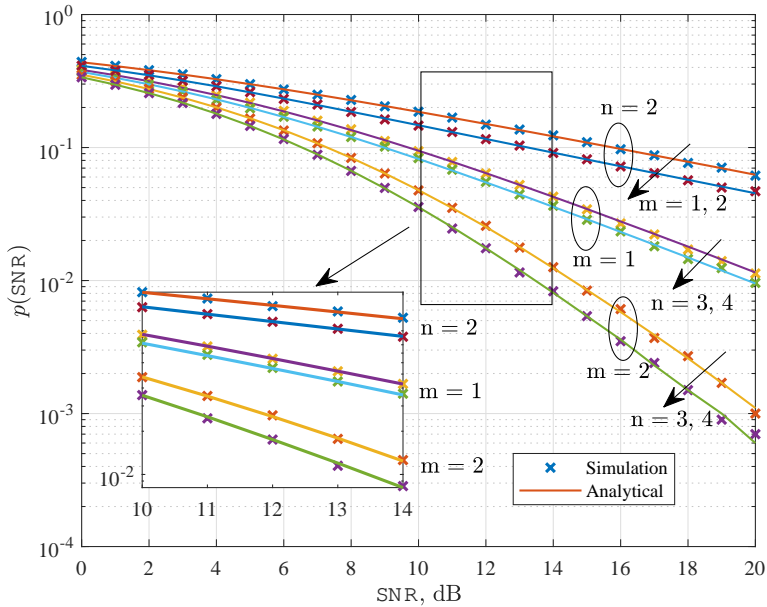}
\caption{Average $\sep$ curves as a function of $\snr$ for QPSK modulation. $n=2,3,4$ and $m=1,2$.}
\label{example1}
\end{figure}

Fig. \ref{example1} plots the average $\sep$ as a function of $\snr$ for QPSK modulation with $n=2,3,4$-bit quantization under Nakagami-$m$ fading with shape parameter $m=1$ and $2$. The simulated results are generated using Monte Carlo simulation, while the analytical results are generated using our expression in \eqref{Eqn:General SEP}. As the plot illustrates, the analytical results accurately follow the simulated results for all cases. We observe a noteworthy improvement in the average $\sep$ when $n$ changes from $2$ to $3$-bit quantization for QPSK modulation in both $m=1$ and $2$. This jump in the average $\sep$ performance is expected in the light of Theorem \ref{Theorem 2}, which states that using one extra bit, on top of $\log_2 M$ bits, improves the $\dvo$ from $\frac12$ to $m$. We also observe that the average $\sep$ reduces as we increase $n$, but the amount by which it reduces also gets smaller as we increase $n$. This can be clearly observed from the zoomed-in section in Fig. \ref{example1}.  As expected, $\dvo=m$ for all $n \geq 3$. Furthermore, $\dvo = \frac{1}{2}$ for any $m$, when $n=2$. 

\begin{figure}[!t]
\center
\includegraphics[width=0.7\textwidth]{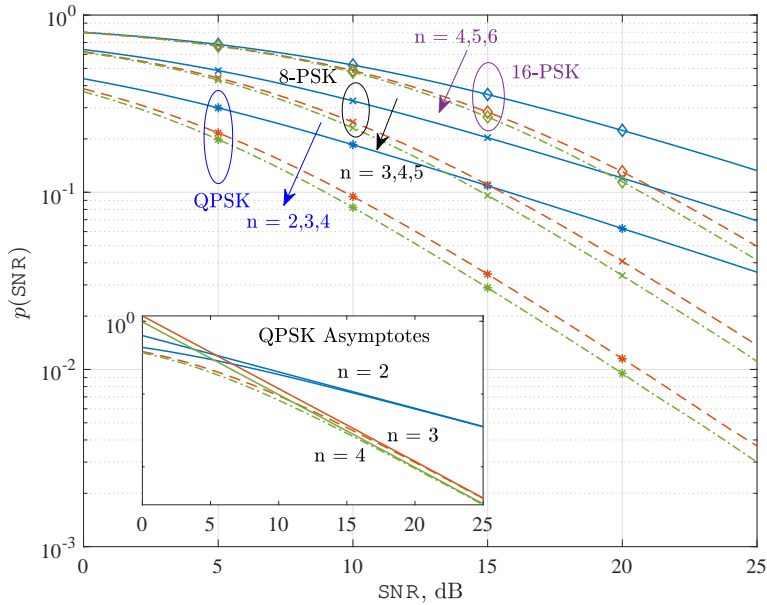}
\caption{Average $\sep$ curves as a function of $\snr$ for different modulation schemes. $n=\log_2M, \log_2M + 1, \log_2M + 2$ and $m=1$.}
\label{example2}
\end{figure}

Fig. \ref{example2} plots the average $\sep$ as a function of $\snr$ for QPSK, $8$-PSK and $16$-PSK modulations schemes while keeping the Nakagami-$m$ shape parameter fixed at $m=1$, which is the classical Rayleigh fading scenario. We plot the average $\sep$ for each modulation scheme by using $n=\log_2M$, $\log_2M+1$ and $\log_2M+2$ bits. From the plots, we can clearly observe that QPSK with $2$-bit, $8$-PSK with $3$-bit and $16$-PSK with $4$-bit quantization have a $\dvo$ of $\frac{1}{2}$. Further, we can observe that QPSK with $3$ or more bits, $8$-PSK with $4$ or more bits, $16$-PSK with 5 or more bits quantizations have a $\dvo$ of $1$, which is equal to $m$ in this case. To further emphasize this point, the zoomed-in section in Fig. \ref{example2} illustrates the asymptotic average $\sep$ versus $\snr$ for QPSK modulation. As stated in Theorem \ref{Theorem 2}, these numerical observations clearly indicate the ternary behaviour in the decay exponent for $p\paren{\snr}$ depending on whether $n \geq \log_2M+1$, $n = \log_2M$, or $n < \log_2M$.

\begin{figure}[!t]
\center
\includegraphics[width=0.7\textwidth]{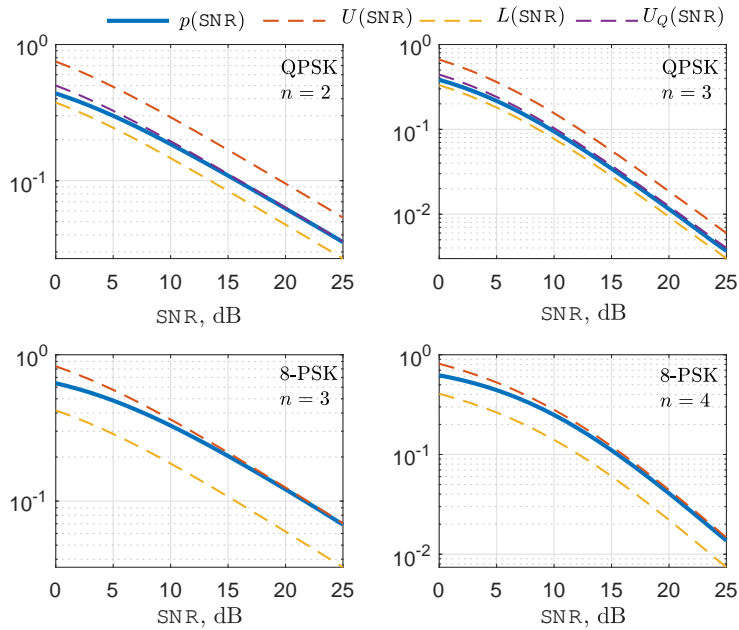}
\caption{Upper and lower bounds on $p\paren{\snr}$ as a function of $\snr$ for QPSK and $8$-PSK modulations. $n=2, 3, 4$ and $m=1$.}
\label{SEP_Bounds}
\end{figure}

In order to illustrate the accuracy of upper and lower bounds on $p\paren{\snr}$, derived in Lemma \ref{Lemma: SEP bounds}, in Fig. \ref{SEP_Bounds} we plot the expressions in \eqref{Eqn: Both bounds}, alongside the exact $p\paren{\snr}$ curve, as a function of $\snr$ for QPSK and $8$-PSK modulations under Rayleigh fading (i.e., $m=1$).  This figure clearly shows that $U\paren{\snr}$ becomes a very tight upper bound for $8$-PSK in the high $\snr$ regime. The figure also confirms that the decay exponents of both $L\paren{\snr}$ and $U\paren{\snr}$ are the same as that of the $p\paren{\snr}$.  

\begin{figure}[!t]
\center
\includegraphics[width=0.7\textwidth]{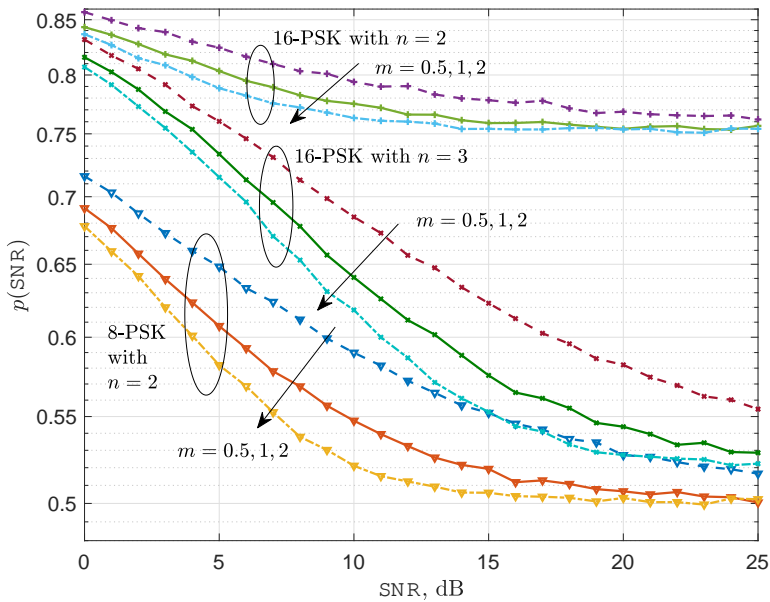}
\caption{Average $\sep$ as a function of $\snr$ for $8$-PSK and $16$-PSK modulations. $n=2<\log_2M$ and $m=0.5,1$ and $2$.}
\label{SEP_less_bits}
\end{figure}

Next, in Fig. \ref{SEP_less_bits}, we plot the simulated average $\sep$ curves as a function of $\snr$ for $8$-PSK modulation with $2$-bit quantization and $16$-PSK modulation with $2$ and $3$-bit quantization. We consider equi-probable transmitted symbols and Nakagami-$m$ fading channel model with $m=0.5,1$ and $2$. The simulated results are again generated by using Monte Carlo simulations. We can clearly observe an error floor for high $\snr$ values when $n < \log_2M$, as established by Theorem \ref{Theorem: lower bound n<log2M}. In particular, the average $\sep$ for $8$-PSK has a lower bound of $0.5$ with $2$-bit quantization. Similarly, the average $\sep$ for $16$-PSK has a lower bound of $0.75$ with $2$-bit quantization and a lower bound of $0.5$ with $3$-bit quantization. It should be noted that the error floor given in Theorem \ref{Theorem: lower bound n<log2M} is more conservative than those observed in Fig. \ref{SEP_less_bits}. This is because it is a universal lower bound that holds for all modulation schemes, quantizer types and fading environments, not only for very specific ones used to plot average $\sep$ curves in Fig. \ref{SEP_less_bits}.

\begin{figure}[!t]
\center
\includegraphics[width=0.7\textwidth]{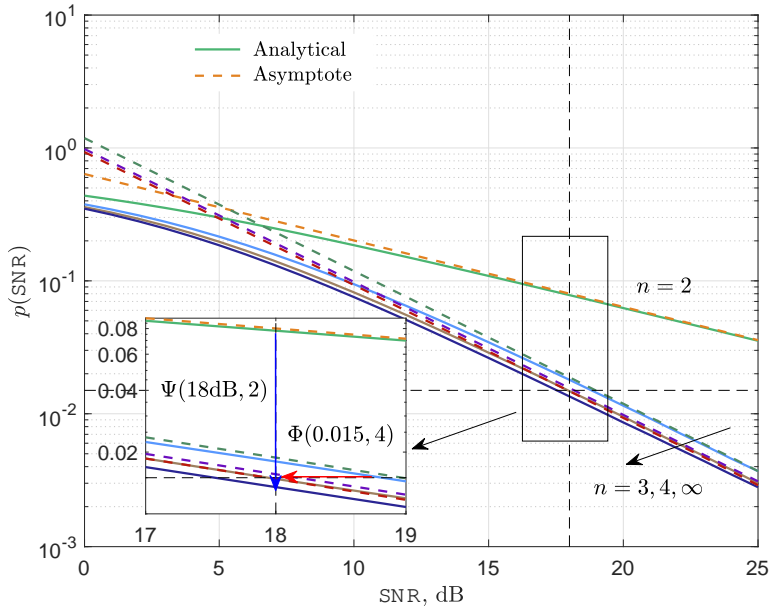}
\caption{Quantization penalty for QPSK modulation and the asymptotic average $\sep$ curves. $n=2,3,4,\infty$ and at $m=1$.}
\label{quantization_penalty}
\end{figure}

Finally, Fig. \ref{quantization_penalty} illustrates the quantization penalty and plots the asymptotic average $\sep$ curves as a function of $\snr$ for QPSK modulation with $n = $ 2, 3, 4 and $\infty$ under Rayleigh fading. The asymptotic plots are generated by using the expressions in \eqref{Eqn: Asymptotic SEP QPSK Rayleigh}. We observe that a $\dvo$ of half is achieved with $2$-bit quantization, and the full $\dvo$ of one is achieved with $n>2$. When the $\snr$ is fixed at $18$ dB, we observe a quantization penalty of $\Psi\paren{18\mbox{ dB},2} \approx 6.35$ dB as we change from $n = 2$ to $\infty$, i.e., we get a $5$-fold increase in the average $\sep$ as we change from $n = 2$ to $\infty$. When the average $\sep$ is fixed at $0.015$, we observe a quantization penalty of $\Phi\paren{0.015,4} \approx 0.8$ dB as we change from $n = 3$ to $4$, i.e., to achieve an average $\sep$ of $0.015$ with $4$-bit quantization, we need $0.8$ dB more transmit power than what is required with $3$-bit quantization.

\section{Conclusions and Future Generalizations} \label{Section: Conclusions}
In this paper, we performed a theoretical analysis of a low-resolution based ADC communication system and obtained fundamental performance limits, optimum ML detectors and a general analytical expression for the average $\sep$ for $M$-PSK modulation with $n$-bit quantization. These results were further investigated for Nakagami-$m$ fading model in detail. We conducted an asymptotic analysis to show that the decay exponent for the average $\sep$ is the same and equal to $m$ with infinite-bit and $n$-bit quantizers for $n \geq \log_2M+ 1$. We also performed an extensive numerical study to illustrate the accuracy of the derived analytical expressions.  

In most parts of the paper, we have focused on phase modulated communications. Phase modulation has an important and practical layering feature enabling the quantizer and detector design separation in low-resolution ADC communications.  For a given number of bits, the quantizer needs to be designed only once, and can be kept constant for all channel realizations. The detector can be implemented digitally as a table look-up procedure using channel knowledge and quantizer output. On the other hand, this feature is lost in joint phase and amplitude modulation schemes such as QAM. The quantizer needs to be dynamically updated for each channel realization in low-resolution ADC based QAM systems. This is because the fading channel amplitude may vary over a wide range, but the phase always varies over $\parenro{-\pi, \pi}$. However, phase modulation is historically known to be optimum only up to modulation order $16$ under peak power limitations \cite{Lucky62}. Hence, it is a notable future research direction to extend the results of this paper to higher order phase and amplitude modulations by taking practical design considerations into account. 

A major result of this paper is the discovery of a ternary $\sep$ behaviour, indicating the sufficiency of $\log_2M+1$ bits for achieving asymptotically optimum $M$-ary communication reliability. Hence, without modifying the conventional RF chain, we can use one extra bit and still achieve the asymptotically optimum communication performance. Another important future research direction is to compare and contrast the backward-compatible receiver design approach of using one extra bit proposed in this paper with other approaches that can potentially modify the conventional RF chain and manipulate the received signals in the waveform domain by introducing extra analog components. This study needs to be done in detail by considering accuracy and agility of analog domain operations, energy consumption of analog and digital circuit components, different modulation schemes and the average $\sep$ performance curves resulting from different low-resolution ADC based receiver architectures. Similarly, utilizing the results of this paper, a further detailed study on the receiver architecture design to determine where to place the diversity combiner (before or after quantizer or detector) and its type is needed when multiple diversity branches are available for data reception.      

\appendices

\section{Proof of Theorem \ref{Theorem: lower bound n<log2M}}
\label{Appendix:Theorem: lower bound n<log2M}
Let us consider a class of hypothetical genie-aided detectors $g:\C^2 \times \sqparen{0:2^{n-1}}  \to \sqparen{0:M-1}$ that has the knowledge of channel noise $W \in \C$, fading coefficient $H \in \C$ and quantizer output $Q\paren{Y} \in \sqparen{0:2^{n-1}}$. We also let $\mathcal{S}_{w, h, k} = \brparen{x \in \mathcal{C}: \sqrt{\snr}hx + w \in \mathcal{R}_k}$ be the set of received signal points resulting in $Q(Y) = k$ for particular realizations of $H=h$ and $W=w$. We first observe that since $n<\log_2M$, there exists at least one quantization region $\mathcal{R}_{\tilde{k}}$ (depending on $w$ and $h$) such that $\mathcal{S}_{w, h, \tilde{k}}$ contains at least $\frac{M}{2^n}$ signal points. We note that $\frac{M}{2^n}$ is always an integer greater than $2$ since $M$ is assumed to be an integer power of $2$. Then, the conditional $\sep$ of any detector $g$ given $W=w$ and $H=h$, which we will denote by $p_g\paren{\snr,h,w}$, can be lower-bounded as 
\begin{align}
p_{g}\paren{\snr,h,w} 
&\geq p_{\min} \sum_{x_i \in \mathcal{S}_{w,h,\tilde{k}}}\PR{g\paren{h,w,\tilde{k}}\neq x_i \, \big| \, W=w, H=h, X=x_i} \nonumber \\
&\geq p_{\min}\paren{\frac{M}{2^n} - 1} \nonumber \\ 
&= \frac{M-2^n}{2^n}p_{\min}, \label{Eqn: Universal SEP Lower Bound 2}
\end{align}

By averaging with respect to $w$ and $h$, we have $p_g\paren{\snr} \geq \frac{M-2^n}{2^n}p_{\min}$, where $p_g\paren{\snr}$ is the average $\sep$ corresponding to detector $g$. This concludes the proof since the obtained lower bound does not depend on the choice of modulation scheme, quantizer structure and detector rule, and hence holds for detectors not utilizing the knowledge of $W$ for any choice of modulation scheme and quantizer structure.

\section{Proof of Theorem \ref{Theorem: ML Detector}}
\label{Appendix: Optimum Detector Proof}
To prove Theorem \ref{Theorem: ML Detector}, we will first obtain the following result.  
\begin{lemma} \label{Lemma: monotonically dec}
Let $\mathcal{R}$ be a convex cone given by $\mathcal{R} = \brparen{z \in \C: \alpha_1 \leq \Arg{z} \leq \alpha_2}$ for $\alpha_1,\alpha_2 \in \parenro{-\pi,\pi}$, and $W_1 \sim \mathcal{CN}\paren{\mu_1,1}$ and $W_2 \sim \mathcal{CN}\paren{\mu_2,1}$ be proper complex Gaussian random variables with means satisfying $\abs{\mu_1} = \abs{\mu_2}=r$ for some $r > 0$. Then, $\PR{W_1 \in  \mathcal{R}} \geq \PR{W_2 \in  \mathcal{R}}$ if $\abs{\mu_1 - z_{\rm mid}} \leq \abs{\mu_2 - z_{\rm mid}}$, where $z_{\rm mid} = r \e{\jmath \frac{\alpha_1 + \alpha_2}{2}}$.
\end{lemma}

\begin{IEEEproof}
It is enough to show this result only for $\alpha_2 = -\alpha_1 = \alpha$. Otherwise, we can first rotate $W_1$, $W_2$ and $\mathcal{R}$ with $\e{-\jmath \frac{\alpha_1 + \alpha_2}{2}}$ and repeat the same calculations below. Let $g\paren{\mu_i} = \PR{W_i \in \mathcal{R}}$ for $i=1,2$, and assume $\abs{\mu_1 - z_{\rm mid}} \leq \abs{\mu_2 - z_{\rm mid}}$. There are multiple cases in which the inequality $\abs{\mu_1 - z_{\rm mid}} \leq \abs{\mu_2 - z_{\rm mid}}$ holds, which we will analyze one-by-one below.       
 
\begin{figure}[!t]
\center
\includegraphics[width=0.7\textwidth]{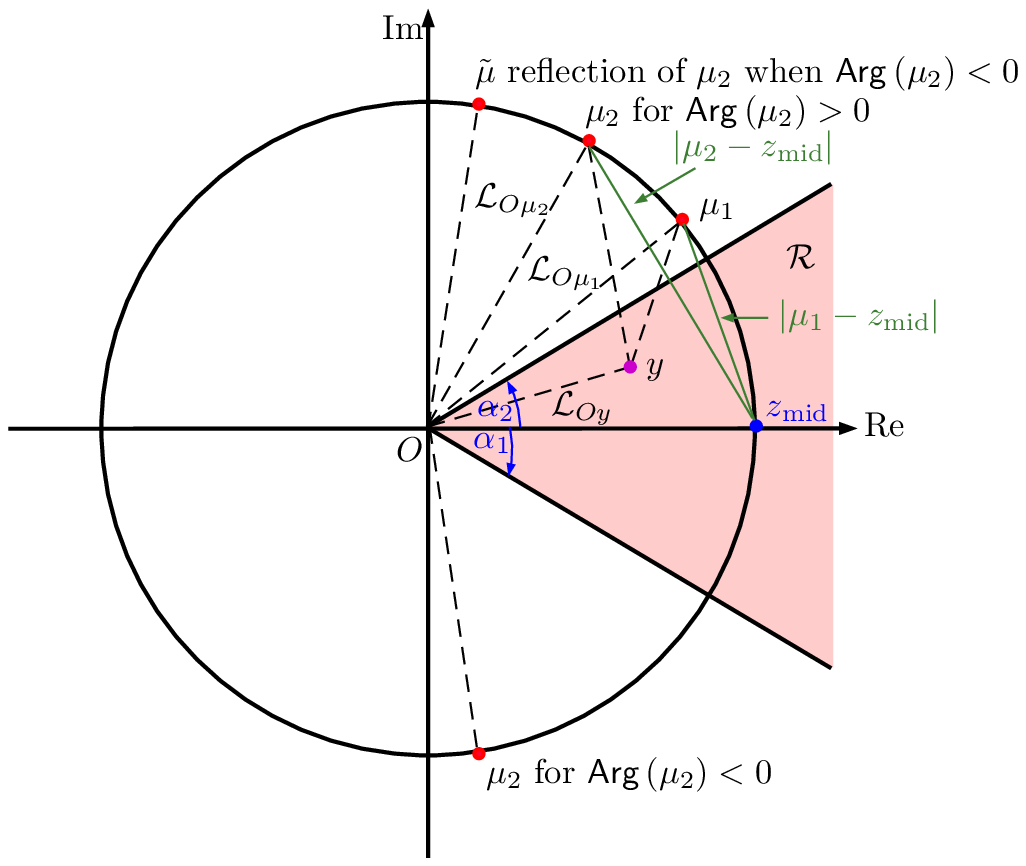}
\caption{An illustration for the proof of Lemma \ref{Lemma: monotonically dec} when $\mu_1$ and $\mu_2$ lie outside $\mathcal{R}^\circ$. $\abs{\mu_1}=\abs{\mu_2} = r$, $\abs{\mu_1 - z_{\rm mid}} \leq \abs{\mu_2 - z_{\rm mid}}$ and $\alpha_2 = -\alpha_1 = \alpha$.}
\label{Lemma1_1}
\end{figure}
 
 First, we will consider the case in which both $\mu_1$ and $\mu_2$ lie outside $\mathcal{R}^\circ$, where $\mathcal{R}^\circ$ is the set of interior points of $\mathcal{R}$. This is the case shown in Fig. \ref{Lemma1_1}.  To start with, we will assume $0 \leq \Arg{\mu_1} \leq \Arg{\mu_2} < \pi$. Then, for any $y \in \mathcal{R}$, the angle between the line segments $\mathcal{L}_{Oy}$ and $\mathcal{L}_{O\mu_1}$ is smaller than the one between the line segments $\mathcal{L}_{Oy}$ and $\mathcal{L}_{O\mu_2}$.\footnote{The line segment $\mathcal{L}_{z_1z_2}$ between the points $z_1 \in \C$ and $z_2 \in \C$ is defined as $\mathcal{L}_{z_1z_2} = \brparen{(1-t) z_1 + t z_2: t \in \sqparen{0,1}}$.} Hence, applying the cosine rule for the triangle formed by $O, y$ and $\mu_1$, and for the triangle formed by $O, y$ and $\mu_2$, it can be seen that $\abs{y - \mu_1} \leq \abs{y - \mu_2}$ for all $y \in \mathcal{R}$.\footnote{This statement is correct even when both $y$ and $\mu_1$ lies on the boundary of $\mathcal{R}$ and the triangle formed by $O, y$ and $\mu_1$ reduces to a line segment.} Therefore, $g\paren{\mu_1} = \frac{1}{\pi}\int_{\mathcal{R}} \exp \paren{-\abs{y-\mu_1}^2 } dy \geq \frac{1}{\pi}\int_{\mathcal{R}} \exp \paren{-\abs{y-\mu_2}^2 } dy = g\paren{\mu_2}$. Next, we assume $\Arg{\mu_2} \in \parenro{-\pi, 0}$ and $0 \leq \Arg{\mu_1} \leq \abs{\Arg{\mu_2}} \leq \pi$. Let $\widetilde{W}$ be the auxiliary random variable distributed according to $\mathcal{CN}\paren{\tilde{\mu} ,1}$ with $\tilde{\mu} = r\e{\jmath \abs{\Arg{\mu_2}}}$, i.e., $\tilde{\mu}$ is the reflection of $\mu_2$ around the real line. Symmetry around the real line implies that $g\paren{\mu_2}$ is equal to $ g\paren{\tilde{\mu}}= \PR{\widetilde{W} \in \mathcal{R}}$, which is less than $g\paren{\mu_1}$ due to our arguments above. For $\Arg{\mu_1} \in \parenro{-\pi, 0}$, the same analysis still holds after reflecting $\mu_1$ around the real line, leading to $g\paren{\mu_1} \geq g\paren{\mu_2}$ for all $\mu_1, \mu_2 \notin \mathcal{R}^\circ$ satisfying $\abs{\mu_1 - z_{\rm mid}} \leq \abs{\mu_2 - z_{\rm mid}}$. 
 
\begin{figure}[!t]
\center
\includegraphics[width=0.7\textwidth]{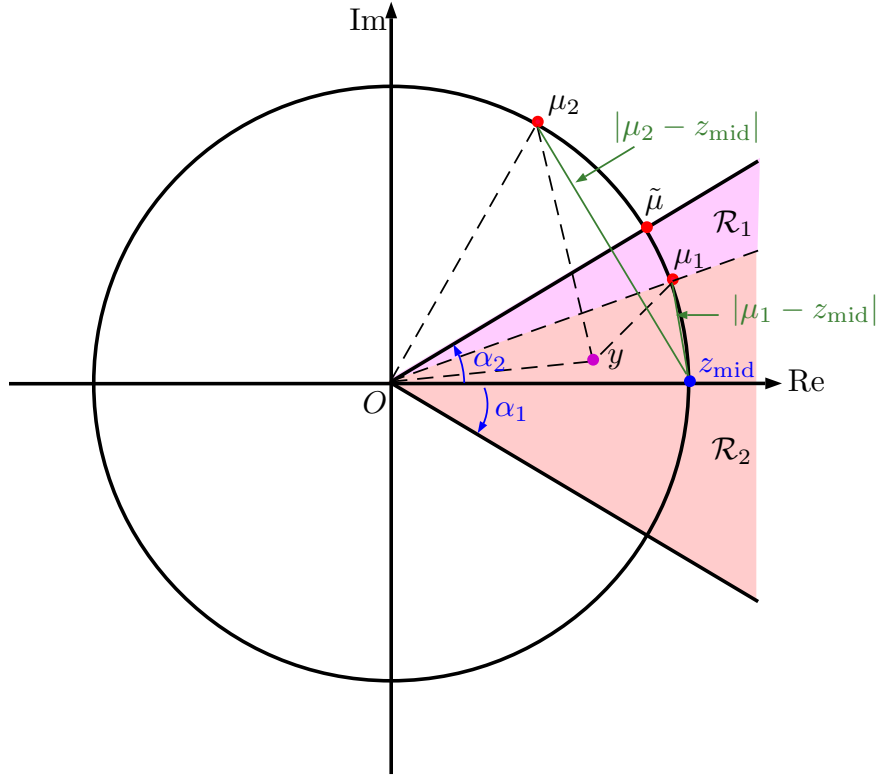}
\caption{An illustration for the proof of Lemma \ref{Lemma: monotonically dec} when $\mu_1 \in \mathcal{R}^\circ$ and $\mu_2 \notin \mathcal{R}^\circ$. $\abs{\mu_1}=\abs{\mu_2} = r$, $\abs{\mu_1 - z_{\rm mid}} \leq \abs{\mu_2 - z_{\rm mid}}$ and $\alpha_2 = -\alpha_1 = \alpha$.}
\label{Lemma1_2}
\end{figure}
 
 Second, we consider the case where $\mu_1 \in \mathcal{R}^\circ$ but $\mu_2 \notin \mathcal{R}^\circ$. This is the case shown in Fig. \ref{Lemma1_2}. It is enough to establish the desired result only for $0 \leq \Arg{\mu_1} \leq \Arg{\mu_2} < \pi$. When $\mu_1$ or $\mu_2$ has a negative phase angle, the same analysis below still holds after reflecting the mean with negative phase around the real line.  Let $\widetilde{W}$ be the auxiliary random variable distributed according to $\mathcal{CN}\paren{\tilde{\mu},1}$ with $\tilde{\mu} = r\e{\jmath \alpha}$, i.e., $\tilde{\mu}$ is located at the upper boundary of $\mathcal{R}$. Our analysis in the first case shows that $g\paren{\tilde{\mu}} = \PR{\widetilde{W} \in \mathcal{R}} \geq g\paren{\mu_2}$ since both $\tilde{\mu}$ and $\mu_2$ are outside $\mathcal{R}^\circ$ and $ 0 \leq \Arg{\tilde{\mu}} \leq \Arg{\mu_2} < \pi$. We next divide $\mathcal{R}$ into two disjoint regions:  $\mathcal{R}_1 = \brparen{z \in \C: \Arg{\mu_1} \leq \Arg{z} \leq \alpha}$ and $\mathcal{R}_2 = \brparen{z \in \C:  -\alpha \leq \Arg{z} < \Arg{\mu_1}}$. Then, we have $\PR{\widetilde{W} \in \mathcal{R}_1} = \PR{W_1 \in \mathcal{R}_1}$ due to symmetry around the line bisecting $\mathcal{R}_1$ and $\PR{\widetilde{W} \in \mathcal{R}_2} \leq \PR{W_1 \in \mathcal{R}_2}$ since $\abs{y-\mu_1} \leq \abs{y-\tilde{\mu}}$ for all $y \in \mathcal{R}_2$. Hence, $g\paren{\mu_1} \geq g\paren{\tilde{\mu}} \geq g\paren{\mu_2}$.  This establishes the desired results for all $\mu_1 \in \mathcal{R}^\circ, \mu_2 \notin \mathcal{R}^\circ$ satisfying $\abs{\mu_1 - z_{\rm mid}} \leq \abs{\mu_2 - z_{\rm mid}}$.
 
 \begin{figure}[!t]
\center
\includegraphics[width=0.7\textwidth]{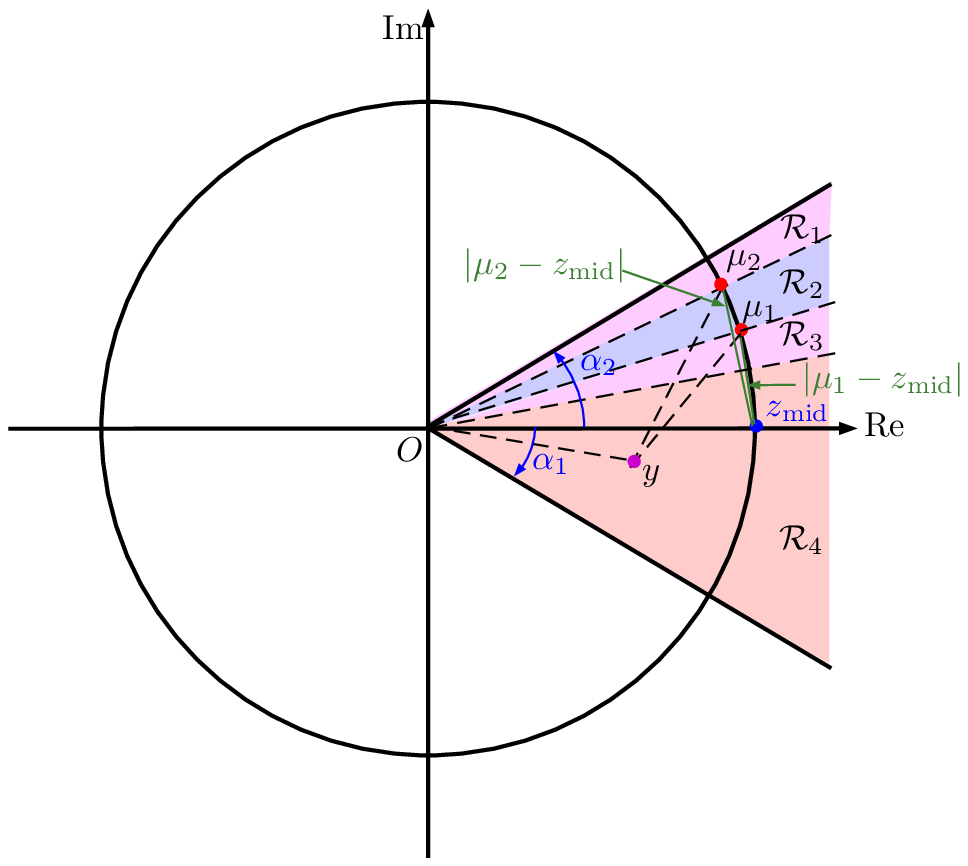}
\caption{An illustration for the proof of Lemma \ref{Lemma: monotonically dec} when $\mu_1$ and $\mu_2$ lie inside $\mathcal{R}^\circ$. $\abs{\mu_1}=\abs{\mu_2} = r$, $\abs{\mu_1 - z_{\rm mid}} \leq \abs{\mu_2 - z_{\rm mid}}$ and $\alpha_2 = -\alpha_1 = \alpha$.}
\label{Lemma1_3}
\end{figure}
 
 Finally, we will consider the third case where both $\mu_1$ and $\mu_2$ lie inside $\mathcal{R}^\circ$. This is the case shown in Fig. \ref{Lemma1_3}. Similar to the first two cases, it is enough to focus only on $0\leq \Arg{\mu_1} \leq \Arg{\mu_2} \leq \alpha$. We divide $\mathcal{R}$ into four disjoint regions given by 
 \begin{eqnarray*}
 \mathcal{R}_1 &=& \brparen{z \in \C: \Arg{\mu_2} \leq \Arg{z} \leq \alpha} \\
 \mathcal{R}_2 &=& \brparen{z \in \C: \Arg{\mu_1} \leq \Arg{z} < \Arg{\mu_2}} \\
 \mathcal{R}_3 &=& \brparen{z \in \C: \Arg{\mu_1} + \Arg{\mu_2} - \alpha \leq \Arg{z} < \Arg{\mu_1}} \\
 \mathcal{R}_4 &=& \brparen{z \in \C:  -\alpha \leq \Arg{z} < \Arg{\mu_1} + \Arg{\mu_2} - \alpha}
 \end{eqnarray*}
 Using the symmetry in the problem, we have $\PR{W_1 \in \mathcal{R}_1} = \PR{W_2 \in \mathcal{R}_3}$, $\PR{W_1 \in \mathcal{R}_2} = \PR{W_2 \in \mathcal{R}_2}$ and $\PR{W_1 \in \mathcal{R}_3} = \PR{W_2 \in \mathcal{R}_1}$. On the other hand, $\PR{W_1 \in \mathcal{R}_4} \geq \PR{W_2 \in \mathcal{R}_4}$ since $\abs{y-\mu_1} \leq \abs{y-\mu_2}$ for all $y \in \mathcal{R}_4$. Hence, $g\paren{\mu_1} \geq g\paren{\mu_2}$ when both $\mu_1$ and $\mu_2$ lie inside $\mathcal{R}^\circ$, which completes the proof.    
 \end{IEEEproof}
 
 \begin{figure}[!t]
\center
\includegraphics[width=0.7\textwidth]{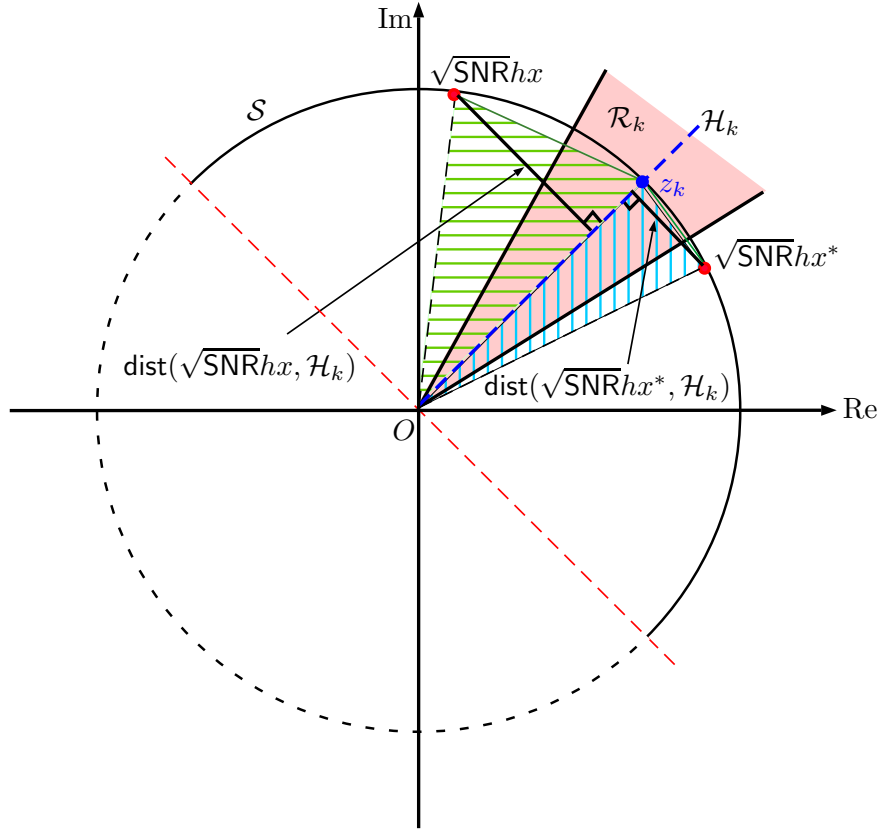}
\caption{An illustration for the proof of Theorem \ref{Theorem: ML Detector} where $\abs{\sqrt{\snr}hx^\star - z_k} \leq \abs{\sqrt{\snr}hx - z_k}$. }
\label{Lemma1_Fig2}
\end{figure}

Now, we will utilize Lemma \ref{Lemma: monotonically dec} to prove Theorem \ref{Theorem: ML Detector}. For $Q\paren{Y} = k$, the ML detector given in \eqref{Eqn: ML detector Q} reduces to finding a signal point in $\mathcal{C}$ maximizing the probability $\PR{\sqrt{\snr}hx + W \in \mathcal{R}_k}$, i.e., 
\begin{align}
    \hat{x}\paren{k, h} \in \underset{x \in \mathcal{C}}{\arg\max }\ \PR{\sqrt{\snr}hx + W \in \mathcal{R}_k}. \nonumber
\end{align}
By Lemma \ref{Lemma: monotonically dec}, $\hat{x}\paren{k,h}$ is the signal point in $\mathcal{C}$ such that $\sqrt{\snr}h \hat{x}\paren{k,h}$ is closest to $z_k = \sqrt{\snr}r\e{\jmath \paren{\frac{2\pi}{2^n}k +\frac{\pi}{2^n}}}$, where $r = \abs{h}$. Further, $\hat{x}\paren{k,h}$ is unique with probability one due to the continuity assumption of the fading distribution. Consider now the semi-circle 
$$\mathcal{S} = \brparen{z \in \C : \abs{z} = \sqrt{\snr}r, \paren{\frac{2\pi}{2^n}k + \frac{\pi}{2^n}-\frac{\pi}{2}} \leq \Arg{z} \leq \paren{\frac{2\pi}{2^n}k + \frac{\pi}{2^n}+\frac{\pi}{2}}}$$
centered around $z_k$ and having $\mathcal{H}_k$ as its bisector. The semi-circle $\mathcal{S}$ is illustrated in Fig \ref{Lemma1_Fig2}. Let $x^\star \in \underset{x \in \mathcal{C}}{\arg\min} \ \dist{\sqrt{\snr}hx, \mathcal{H}_k}$.  For the $M$-PSK modulation scheme ($M\geq 2$) with regularly spaced signal points on the unit circle, we always have $\sqrt{\snr} h x^\star \in \mathcal{S}$ and $\sqrt{\snr} h \hat{x}(k,h) \in \mathcal{S}$. Take now another signal point $x \in \mathcal{C}$ different than $x^\star$ and satisfying $\sqrt{\snr} h x \in \mathcal{S}$. Consider the triangle formed by $0, z_k$ and $\sqrt{\snr} h x^\star$, and the one formed by $0, z_k$ and $\sqrt{\snr} h x$. We first observe that the area of the first triangle is smaller than the area of the second one since they share the line segment $\mathcal{L}_{Oz_k}$ as their common base but the height of the first one $\dist{\sqrt{\snr}hx^\star, \mathcal{H}_k}$ corresponding to this base is smaller than the height of the second one $\dist{\sqrt{\snr}hx, \mathcal{H}_k}$ corresponding to the same base. This is also illustrated in Fig. \ref{Lemma1_Fig2}. This observation, in turn, implies $\abs{\sqrt{\snr}hx^\star - z_k} \leq \abs{\sqrt{\snr}hx - z_k}$ because the remaining side lengths of both triangles are equal to $\sqrt{\snr} r$. Since this is correct for any $x \in \mathcal{C}$ satisfying $\sqrt{\snr} h x \in \mathcal{S}$, we conclude that $x^\star$ is unique and equal to $x^\star = \hat{x}(k, h)$.                 

\section{Proof of Lemma \ref{Lemma: p(snr) = 2^n P_00}} \label{Appendix:p(snr)=2^nP00 Proof}
The proof of Lemma \ref{Lemma: p(snr) = 2^n P_00} is based on an application of the law of total probability \cite{Tsitsiklis08}. To this end, we consider a partition $\brparen{\mathcal{D}_k}_{k=0}^{2^n-1}$ of $\C$, where each element of this partition is given by $\mathcal{D}_k = \brparen{z \in \C: \paren{2k -1}\frac{\pi}{2^n} \leq \Arg{z} +\pi < \paren{2k+1}\frac{\pi}{2^n}}$ for $k \in \sqparen{1:2^n-1}$, and $\mathcal{D}_0 = \brparen{z \in \C: \pi - \frac{\pi}{2^n} \leq \Arg{z} < \pi} \bigcup \brparen{z \in \C: -\pi \leq \Arg{z} < \frac{\pi}{2^n} - \pi}$.  Let $x_i = \e{\jmath \pi\paren{\frac{2i+1}{M} - 1}}$ be the $i$th signal point in the constellation set $\mathcal{C}$ for $i \in \sqparen{0:M-1}$. Then, we can express $p\paren{\snr}$ according to    
\begin{align}
p\paren{\snr} 
&= \frac{1}{M} \sum_{i=0}^{M-1}\sum_{k=0}^{2^n-1} \int\limits_{\mathcal{D}_k}\PR{x_i \neq \hat{x}\paren{Q\paren{Y},h}|H=h, X=x_i} f_H\paren{h} dh. \label{Eqn: p(SNR) 2}
\end{align}

We will show that all the terms in \eqref{Eqn: p(SNR) 2} are equal to each other. Next, we define $\mathcal{E}_i = \brparen{z \in \C: \Arg{x_i} - \frac{\pi}{M} \leq \Arg{z} < \Arg{x_i} + \frac{\pi}{M}}$ for $i \in \sqparen{0: M-1}$. Note that $\mathcal{E}_i$ contains all $\mathcal{H}_k$'s (i.e., bisectors of quantization regions) to which $x_i$ is the closest signal point since $x_i$'s are uniformly spaced on the unit circle in $\C$. Furthermore, this statement continues to be true for $\sqrt{\snr}hx_i$ as long as $\Arg{h} \in \parenro{-\frac{\pi}{2^n}, \frac{\pi}{2^n}}$ since the angular spacing between $\mathcal{H}_k$'s is uniform and equal to $\frac{2\pi}{2^{n}}$. Notice that $\Arg{h} \in \parenro{-\frac{\pi}{2^n}, \frac{\pi}{2^n}}$ if and only if $h \in \mathcal{D}_{2^{n-1}}$. On the other hand, if $\Arg{h} \in \parenro{\frac{\pi}{2^n}, \frac{3 \pi}{2^n}}$, the region $\e{\jmath \frac{2\pi}{2^{n}}}\mathcal{E}_i = \brparen{\e{\jmath \frac{2\pi}{2^{n}}} z \in \C: \Arg{x_i} - \frac{\pi}{M} \leq \Arg{z} < \Arg{x_i} + \frac{\pi}{M}}$ contains all $\mathcal{H}_k$'s to which $\sqrt{\snr}hx_i$ is the closest. Notice also that $\Arg{h} \in \parenro{\frac{\pi}{2^n}, \frac{3 \pi}{2^n}}$ if and only if $h \in \mathcal{D}_{2^{n-1}+1}$. Similarly, $\e{-\jmath \frac{2\pi}{2^{n}}}\mathcal{E}_i$ contains all $\mathcal{H}_k$'s to which $\sqrt{\snr}hx_i$ is closest if $\Arg{h} \in \parenro{-\frac{3\pi}{2^n}, -\frac{\pi}{2^n}}$, and $\Arg{h} \in \parenro{-\frac{3\pi}{2^n}, -\frac{\pi}{2^n}}$ if and only if $h \in \mathcal{D}_{2^{n-1}-1}$. The same idea extends to any $\mathcal{D}_k$, and we define
\begin{eqnarray}
\mathcal{E}_{i,k} \defeq \exp\paren{\jmath\paren{k-2^{n-1}}\frac{2\pi}{2^n}} \mathcal{E}_i, \hspace{0.15cm} \mbox{ for } i \in \sqparen{0:M-1} \mbox{ and } k \in \sqparen{0:2^n-1}.  \label{Eqn: Equal Probability Sets}
\end{eqnarray}
We will use the sets defined in \eqref{Eqn: Equal Probability Sets} to show that all the terms in \eqref{Eqn: p(SNR) 2} are equal.     

To complete the proof, we let $p_{i,k} = \int_{\mathcal{D}_k}\PR{x_i \neq \hat{x}\paren{Q\paren{Y},h}|H=h, X=x_i} f_H\paren{h} dh$ for $i \in \sqparen{0:M-1}$ and $k \in \sqparen{0:2^n-1}$. We also define $\theta^{\prime}_i = -\pi\paren{\frac{2i}{M}-1}$, $\theta^{\prime\prime}_k = -\paren{k - 2^{n-1}}\frac{2\pi}{2^n}$, and $\theta_{i,k} = \theta^{\prime}_i + \theta^{\prime\prime}_k$ for $i \in \sqparen{0:M-1}$ and $k \in \sqparen{0:2^n-1}$. We first observe that $\e{\jmath \theta_{i, k}} \mathcal{E}_{i, k} = \mathcal{E}_{\frac{M}{2}}$ since multiplication with $\e{\jmath \theta^{\prime}_i}$ rotates the $i$th signal point to $x_{\frac{M}{2}}$ and multiplication with $\e{\jmath \theta^{\prime\prime}_k}$ removes the effect of partition selection for $h$. Secondly, we observe that when $h \in \mathcal{D}_k$, the event $\brparen{x_i \neq \hat{x}\paren{Q\paren{Y},h}}$ is equivalent to $\brparen{Y \notin \mathcal{E}_{i, k}}$  since $\mathcal{E}_{i, k}$ contains all bisectors to which $\sqrt{\snr}h x_i$ is closest for this range of $h$ values. Hence, the following chain of equalities hold:
\begin{eqnarray}
p_{i,k} &\stackrel{\rm (a)}{=}& \int_{\mathcal{D}_k}\PR{\sqrt{\snr}h x_i + W \notin \mathcal{E}_{i,k}}f_H\paren{h} dh \nonumber \\
&=& \int_{\mathcal{D}_k}\PR{ W \notin \mathcal{E}_{i,k} - \sqrt{\snr}h x_i}f_H\paren{h} dh \nonumber \\
&=& \int_{\mathcal{D}_k} \PR{ \e{\jmath \theta_{i, k}}W \notin \e{\jmath \theta_{i, k}} \mathcal{E}_{i,k} - \e{\jmath \theta_{i, k}}\sqrt{\snr}h x_i}f_H\paren{h} dh \nonumber \\
&\stackrel{\rm (b)}{=}& \int_{\mathcal{D}_k} \PR{ W \notin \mathcal{E}_{\frac{M}{2}} - \sqrt{\snr}\e{\jmath \theta^{\prime\prime}_{k}}h x_\frac{M}{2}}f_H\paren{h} dh, \label{Eqn: pik Derivation 1} 
\end{eqnarray}
where (a) follows from the independence of $W$, $H$ and $X$, and (b) follows from above observations and the circular symmetry property of $W$. Let us now define $z = \e{\jmath \theta^{\prime\prime}_k} h$ in \eqref{Eqn: pik Derivation 1}. Since multiplication with a unit magnitude complex number is a unitary transformation (i.e., rotation) over the complex plane, we have   
\begin{eqnarray}
p_{i,k} &=& \int_{\mathcal{D}_k} \PR{ W \notin \mathcal{E}_{\frac{M}{2}} - \sqrt{\snr}\e{\jmath \theta^{\prime\prime}_{k}}h x_\frac{M}{2}}f_H\paren{h} dh \nonumber \\
&=& \int_{\e{\jmath \theta^{\prime\prime}_{k}} \mathcal{D}_k} \PR{ W \notin \mathcal{E}_{\frac{M}{2}} - \sqrt{\snr}z x_\frac{M}{2}}f_H\paren{\e{-\jmath \theta^{\prime\prime}_{k}} z} dz \nonumber \\ 
&\stackrel{\rm (a)}{=}& \int_{\mathcal{D}_{2^{n-1}}} \PR{ W \notin \mathcal{E}_{\frac{M}{2}} - \sqrt{\snr}z x_\frac{M}{2}}f_H\paren{z} dz \nonumber \\ 
&\stackrel{\rm (b)}{=}&  \int_{\mathcal{D}_{2^{n-1}}} \PR{ \sqrt{\snr}z x_\frac{M}{2} + W \notin \mathcal{E}_{\frac{M}{2}, 2^{n-1}}}f_H\paren{z} dz \nonumber \\
&\stackrel{\rm (c)}{=}& p_{\frac{M}{2}, 2^{n-1}},
\end{eqnarray}
where (a), (b) and (c) follow from the circular symmetry of $H$ \cite{Picinbono94, Koivunen12} and the corresponding definitions for $\mathcal{D}_k$, $\mathcal{E}_{i,k}$ and $p_{i,k}$ for $i \in \sqparen{0:M-1}$ and $k \in \sqparen{0:2^n-1}$. This shows $p\paren{\snr} = 2^n p_{\frac{M}{2}, 2^{n-1}}$. For a circularly-symmetric pdf $f_H\paren{h}$, it is well-known that $rf_H\paren{r\cos \lambda, r\sin \lambda} = \frac{1}{2\pi}f_R\paren{r}$ \cite{book_8}. Switching to polar coordinates, and using the identities $rf_H\paren{r\cos \lambda, r\sin \lambda} = \frac{1}{2\pi}f_R\paren{r}$, $x_\frac{M}{2} = \e{\jmath\frac{\pi}{M}}$ and $\mathcal{E}_{\frac{M}{2}, 2^{n-1}}=\mathcal{E}$, we have  
\begin{align}
p(\snr) &= \frac{2^{n-1}}{\pi} \int_{-\frac{\pi}{2^n}}^{\frac{\pi}{2^n}}\int_{0}^{\infty}\PR{\sqrt{\snr}r \e{\jmath\paren{\frac{\pi}{M} +\lambda}} + W \notin \mathcal{E}}f_{R}\paren{r}\,dr \, d\lambda. \nonumber 
\end{align}
By using the change of variables $\theta = \frac{\pi}{M} + \lambda$, we conclude the proof.

\section{Proof of Lemma \ref{Lemma: sum of pi(SNR) limit}}
\label{Appendix: Proof of Lemma: sum of pi(SNR) limit}
The proof of part (i) follows immediately from Definition \ref{Def: Exponential Equality}:
\begin{eqnarray}
\lim_{\snr \ra \infty} \frac{\log\paren{\alpha f\paren{\snr}}}{\log \snr} = \lim_{\snr \ra \infty} \frac{\log\alpha}{\log\snr} + \lim_{\snr \ra \infty} \frac{\log f\paren{\snr}}{\log \snr} = d_i. \nonumber  
\end{eqnarray}
For the proof of part (ii), given any $\epsilon > 0$, let $c>0$ be such that  
\begin{eqnarray}
\snr^{d_i-\epsilon} \leq f_i\paren{\snr} \leq \snr^{d_i + \epsilon}
\end{eqnarray}
for all $\snr \geq c$ and $i \in \sqparen{1:N}$. Let $i^\prime = \arg\max_{i \in \sqparen{1:N}} d_i$. Then, as $\snr \ra \infty$, we can write 
\begin{eqnarray}
\log\paren{\sum_{i=1}^N f_i\paren{\snr}} &\leq& \log\paren{\sum_{i=1}^N \snr^{d_i + \epsilon}} \nonumber \\
&=& \log \snr^{d_{\max} +\epsilon} + \log\paren{1 + \sum_{\genfrac{}{}{0pt}{}{i=1}{i \neq i^\prime}}^N\snr^{d_i - d_{\max}}} \nonumber \\
&=& \paren{d_{\max} + \epsilon} \log\snr + \LO{1}, \nonumber
\end{eqnarray}
which implies $\sum_{i=1}^N f_i\paren{\snr} \eleq d_{\max} + \epsilon$. Since $\epsilon > 0$ is arbitrary, we conclude that $\sum_{i=1}^N f_i\paren{\snr} \eleq d_{\max}$. The other direction $\sum_{i=1}^N f_i\paren{\snr} \egeq d_{\max}$ follows from the same arguments, which completes the proof.     

\section{Proof of Lemma \ref{Lemma: p1(SNR) limit}}
\label{Appendix:Proof of Lemma: p1(SNR) limit}
We start with the case $M=4$ and $n=2$, and obtain upper and lower bounds on $p_1\paren{\snr}$ that will lead to the same exponential equality. For the upper bound, we write $p_1\paren{\snr}$ as  
\begin{align}
p_1\paren{\snr} &= \frac{2m^m}{\pi^{2}}\int_{0}^{\frac{\pi}{2}}\int_{0}^{\frac{\pi}{2}} \left( \frac{\snr}{\sin^2\beta}\cos^2\theta + m \right)^{-m}  d\theta d\beta \nonumber \\
&\leq \frac{m^m}{\pi}\int_{0}^{\frac{\pi}{2}} \left( \snr\cos^2\theta + m \right)^{-m}  d\theta. \nonumber
\end{align}
Let $\theta^*\paren{\snr}$ be such that $\cos\paren{\theta^*\paren{\snr}}= \snr^{-\frac12}$. Then, $\theta^*\paren{\snr} = \arccos\paren{\snr^{-\frac12}} = \frac{\pi}{2}-\snr^{-\frac12} - \LO{\snr^{-\frac12}}$ as $\snr \to \infty$. 
Using the fact that the line $1-\frac{2}{\pi}\theta$ is a lower bound for $\cos\theta$ for $\theta \in \sqparen{0, \frac{\pi}{2}}$, we have
\begin{align}
p_1\paren{\snr} &\leq \frac{m^m}{\pi}\int_{0}^{\theta^*\paren{\snr}} \paren{ \snr\cos^2\theta + m }^{-m}  d\theta +\frac{m^m}{\pi}\int_{\theta^*\paren{\snr}}^{\frac{\pi}{2}} \paren{ \snr\cos^2\theta + m }^{-m}  d\theta \nonumber \\
&\leq \frac{m^m}{\pi}\snr^{-m}\int_{0}^{\theta^*\paren{\snr}} \paren{ 1-\frac{2}{\pi}\theta}^{-2m}  d\theta +\frac{1}{\pi} \snr^{-\frac12}\paren{1 + \LO{1}} \nonumber \\ 
&= - \frac{m^m}{2}\snr^{-m} \int_1^{1-\frac{2}{\pi}\theta^*\paren{\snr}} u^{-2m} du +\frac{1}{\pi} \snr^{-\frac12}\paren{1 + \LO{1}} \nonumber \\
&= \left \{ \begin{array}{ll}
            \snr^{-\frac{1}{2}} \BO{\log{\snr}} & \mbox{ if } m = \frac12 \\
            \snr^{-\frac{1}{2}}\BO{1} & \mbox{ if } m > \frac12 \end{array} \right. \label{Eqn: upper bound for p1} 
\end{align}
for large values of $\snr$. Equation \eqref{Eqn: upper bound for p1} shows that $p_1\paren{\snr} \eleq \snr^{-\frac12}$ for all $m\geq \frac12$ when $M=4$ and $n=2$.  For the other direction, we obtain a lower bound on $p_1\paren{\snr}$ as below.  
\begin{align}
p_1\paren{\snr} &= \frac{2m^m}{\pi^{2}}\int_{0}^{\frac{\pi}{2}}\int_{0}^{\frac{\pi}{2}} \paren{\sin\beta}^{2m} \paren{\snr\cos^2\theta + m \sin^2\beta }^{-m}  d\theta d\beta \nonumber \\
&\geq \frac{2m^m}{\pi^{2}}\int_{0}^{\frac{\pi}{2}} \paren{\sin\beta}^{2m} d\beta \int_{0}^{\frac{\pi}{2}} \paren{\snr\cos^2\theta + m }^{-m}  d\theta \nonumber
\end{align}
We observe that $\int_{0}^{\frac{\pi}{2}} \paren{\sin\beta}^{2m} d\beta = \frac{\sqrt{\pi}\Gamma\paren{m + \frac{1}{2}}}{2\Gamma\paren{m+1}}$ for $m>-\frac{1}{2}$ and let $c = \frac{m^m}{\pi^{1.5}} \frac{\Gamma\paren{m + \frac{1}{2}}}{\Gamma\paren{m+1}}$. We first consider the case $m=\frac12$.  Then, as $\snr \ra \infty$, we have
\begin{eqnarray}
p_1\paren{\snr} &\geq& c \snr^{-\frac12} \int_0^{\frac{\pi}{2}} \paren{\cos^2\theta +1}^{-\frac12} d\theta \nonumber \\ 
&=& \snr^{-\frac12} \OO{1}. \label{Eqn: p1 Lower Bound 1}
\end{eqnarray}

For $m>\frac12$, we define $\theta^*\paren{\snr}$ as above and lower bound $p_1\paren{\snr}$ for large values of $\snr$ as
\begin{eqnarray}
p_1\paren{\snr} &\geq& c \snr^{-m} \int_{\theta^*\paren{\snr}}^{\frac{\pi}{2}} \paren{\cos^2\theta + \frac{m}{\snr}}^{-m} d\theta \nonumber \\ 
&\geq& c \snr^{-m} \int_{\theta^*\paren{\snr}}^{\frac{\pi}{2}} \paren{\frac{1}{\snr} + \frac{m}{\snr}}^{-m} d\theta \nonumber \\ 
&=& \snr^{-\frac12} c (1+m)^{-m} \paren{1 + \LO{1}} \nonumber \\
&=& \snr^{-\frac12} \OO{1}. \label{Eqn: p1 Lower Bound 2}
\end{eqnarray}

Using \eqref{Eqn: p1 Lower Bound 1} and \eqref{Eqn: p1 Lower Bound 2}, we conclude that $p_1\paren{\snr} \egeq \snr^{-\frac12}$ for all $m\geq \frac12$ when $M=4$ and $n=2$. Since $p_1\paren{\snr}$ also satisfies $p_1\paren{\snr} \eleq \snr^{-\frac12}$ in this case, we have $p_1\paren{\snr} \eeq \snr^{-\frac12}$ for $m\geq \frac12$, $M=4$ and $n=2$. 

Next, we consider $M=4$ and $n>2$. In this case, we write $p_1\paren{\snr}$ as
\begin{eqnarray*}
p_1\paren{\snr} = \frac{2^{n-1}m^m}{\pi^2}\snr^{-m} \int_{0}^{\frac{\pi}{2}}\int_{\frac{\pi}{4} - \frac{\pi}{2^n}}^{\frac{\pi}{4} + \frac{\pi}{2^n}} \paren{ \frac{\cos^2\theta}{\sin^2\beta} + \frac{m}{\snr} }^{-m}  d\theta d\beta.
\end{eqnarray*}
Let $g_{\snr}\paren{\theta, \beta} = \paren{\frac{\cos^2\theta}{\sin^2\beta} + \frac{m}{\snr}}^{-m}$ be a collection of functions indexed by $\snr$. These functions increase to $g_\infty\paren{\theta, \beta} = \paren{\frac{\cos\theta}{\sin\beta}}^{-2m}$ as $\snr \ra \infty$. Further, $\int_{0}^{\frac{\pi}{2}}\int_{\frac{\pi}{4} - \frac{\pi}{2^n}}^{\frac{\pi}{4} + \frac{\pi}{2^n}} g_\infty\paren{\theta, \beta} d\theta d\beta < \infty$. Hence, as $\snr \ra \infty$, we conclude that 
\begin{eqnarray*}
p_1\paren{\snr} = \snr^{-m} \TO{1}
\end{eqnarray*}
by using the monotone convergence theorem \cite{book_7}, which implies $p_1\paren{\snr} \eeq \snr^{-m}$. The proof for $M>4$ and $n\geq \log_2M$ is similar, and we omit it to avoid repetition.      

\section{Proof of Lemma \ref{Lemma: p2(SNR) limit}}
\label{Appendix:Proof of Lemma: p2(SNR) limit}
For $M=4$ and $n=2$, $p_2\paren{\snr}=p_1\paren{\snr}$, and hence the proof of $p_2\paren{\snr} \eeq \snr^{-\frac12}$ in this case directly follows from Lemma \ref{Lemma: p1(SNR) limit}. For $M>4$ and $n > \log_2M$, a similar argument using the monotone convergence theorem as in the proof of Lemma \ref{Lemma: p1(SNR) limit} readily shows that $p_2\paren{\snr} \eeq \snr^{-m}$ for all $m \geq \frac12$. Therefore, we will only focus on the case $M>4$ and $n=\log_2M$ to complete the proof of Lemma \ref{Lemma: p2(SNR) limit}.   

For $M>4$ and $n=\log_2M$, we will obtain upper and lower bounds on $p_2\paren{\snr}$ leading to the same exponential equality. For the upper bound, we have
\begin{eqnarray}
p_2\paren{\snr} &=& \frac{2^{n-1}m^m}{\pi^2}\int_0^{\frac{\pi}{2}}\int_0^{\frac{2\pi}{M}} \paren{\frac{\snr}{\sin^2\beta} \sin^2\theta + m}^{-m}d\theta d\beta \nonumber \\
&\leq& \frac{2^{n-1}m^m}{\pi^2}\int_0^{\frac{\pi}{2}}\int_0^{\frac{\pi}{2}} \paren{\frac{\snr}{\sin^2\beta} \sin^2\theta + m}^{-m}d\theta d\beta \nonumber \\
&\eeq& \snr^{-\frac12}, \label{Eqn: p2(SNR) exp upper bound}
\end{eqnarray}
where the last equality follows from the fact that $p_2\paren{\snr} \eeq \snr^{-\frac12}$ for $M=4$ and $n=2$ and Lemma \ref{Lemma: sum of pi(SNR) limit}.

For the lower bound, we have
\begin{eqnarray}
p_2\paren{\snr} &=& \frac{2^{n-1}m^m}{\pi^2}\int_0^{\frac{\pi}{2}}\int_0^{\frac{2\pi}{M}} \paren{\sin\beta}^{2m}\paren{\snr \sin^2\theta + m\sin^2\beta}^{-m}d\theta d\beta \nonumber \\
&\geq& c \snr^{-m}\int_0^{\frac{2\pi}{M}} \paren{\sin^2\theta + \frac{m}{\snr}}^{-m}d\theta, \nonumber
\end{eqnarray}
where $c=\frac{2^{n-2}m^m}{\pi^{1.5}}\frac{\Gamma\paren{m+\frac12}}{\Gamma\paren{m+1}}$. Let $\theta^*\paren{\snr}$ be such that $\sin\paren{\theta^*\paren{\snr}} = \snr^{-\frac12}$. Then, $\theta^*\paren{\snr} = \arcsin\paren{\snr^{- \frac12}} = \snr^{-\frac12} + \LO{\snr^{-\frac12}}$ as $\snr \ra \infty$. Hence, as $\snr \ra \infty$, we have   
\begin{eqnarray}
p_2\paren{\snr} &\geq& c \snr^{-m}\int_0^{\theta^*\paren{\snr}} \paren{\sin^2\theta + \frac{m}{\snr}}^{-m}d\theta \nonumber \\
&\geq& c \snr^{-m} \int_0^{\theta^*\paren{\snr}} \paren{\frac{1}{\snr} + \frac{m}{\snr}}^{-m} d\theta \nonumber \\
&=& \snr^{-\frac12} \OO{1}, \label{Eqn: p2(SNR) exp lower bound}
\end{eqnarray}
which implies $p_2\paren{\snr} \egeq \snr^{-\frac12}$. Since $p_2\paren{\snr}$ also satisfies $p_2\paren{\snr} \eleq \snr^{-\frac{1}{2}}$ in this case, we have $p_2\paren{\snr} \eeq \snr^{-\frac{1}{2}}$ for $m \geq \frac{1}{2}$, $M >4$ and $n=\log_2M$.

\section{Proof of Lemma \ref{Lemma: SEP bounds}}
\label{Appendix: Proof of Lemma: SEP bounds}
We will prove this lemma for general circularly-symmetric fading processes. To this end, let $H = R \e{\jmath \Theta}$ be the circularly-symmetric fading coefficient with the joint phase and magnitude pdf $f_{R, \Theta}\paren{r, \theta} = \frac{1}{2\pi} f_R\paren{r}$ for $\theta \in \parenro{-\pi, \pi}$ and $r \geq 0$. In the proof of Theorem \ref{Theorem: General SEP}, we obtained the expression $p\paren{\snr, h}$ with $h = r\e{\jmath\theta}$ given by
\begin{align}
p\paren{\snr, h} \nonumber \\
&\hspace{-1.5cm}= \mathcal{Q}\paren{\sqrt{2\snr}r\cos\theta} + \mathcal{Q}\paren{\sqrt{2\snr}r\sin\theta} - \mathcal{Q}\paren{\sqrt{2\snr}r\cos\theta}\mathcal{Q}\paren{\sqrt{2\snr}r\sin\theta} \hspace{3cm} \nonumber \\
&\hspace{-1.5cm}+  \frac{1}{\sqrt{\pi}} \int_{-\sqrt{\snr}r\cos\theta}^\infty \mathcal{Q}\paren{\sqrt{2\snr}r\sec\paren{\frac{2\pi}{M}}\sin\paren{\frac{2\pi}{M} - \theta} + \sqrt{2}\tan\paren{\frac{2\pi}{M}}w}\e{-w^2} dw .\label{Eqn: p(SNR, h) Expansion}
\end{align}
Below, we will always use $r\e{\jmath \theta}$ as the polar coordinate representation of $h$, i.e., $r = \abs{h}$ and $\theta = \Arg{h}$. Integrating $2^n p\paren{\snr, h}$ with respect to  $f_{R, \Theta}\paren{r, \theta}$ for $\theta \in \parenro{\frac{\pi}{M} - \frac{\pi}{2^n}, \frac{\pi}{M} + \frac{\pi}{2^n}}$ and $r \geq 0$, and using the Nakagami-$m$ pdf for $f_{R}(r)$ together with Craig's formula, we obtained the resulting $p\paren{\snr}$ expression in Theorem \ref{Theorem: General SEP}. Here, we will not assume any specific functional form for $f_{R}(r)$.     

We start with obtaining the lower bound $L\paren{\snr}$ on $p\paren{\snr}$. Let 
\begin{eqnarray}
p_1\paren{\snr, h} &=& \mathcal{Q}\paren{\sqrt{2\snr}r\cos\theta} \nonumber \\
p_2\paren{\snr, h} &=& \mathcal{Q}\paren{\sqrt{2\snr}r\sin\theta} \nonumber \\
p_3\paren{\snr, h} &=& \mathcal{Q}\paren{\sqrt{2\snr}r\cos\theta}\mathcal{Q}\paren{\sqrt{2\snr}r\sin\theta} \nonumber 
\end{eqnarray}
and $p_4\paren{\snr, h}$ be the last integral term in \eqref{Eqn: p(SNR, h) Expansion}. For $i \in \sqparen{1:4}$, $p_i\paren{\snr}$ is defined to be the integral of $2^n p_i\paren{\snr, h}$  with respect to $f_{R, \Theta}\paren{r, \theta}$ for $\theta \in \parenro{\frac{\pi}{M} - \frac{\pi}{2^n}, \frac{\pi}{M} + \frac{\pi}{2^n}}$ and $r \geq 0$. For the given integration range, $p_3\paren{\snr, h} \leq \frac12 p_2\paren{\snr, h}$ since the argument of the $\mathcal{Q}$-function is always positive. Hence, we have
\begin{eqnarray}
p\paren{\snr, h} &\geq& p_1\paren{\snr, h} + p_2\paren{\snr, h} - p_3\paren{\snr, h} \nonumber \\
&\geq& p_1\paren{\snr, h} + \frac12 p_2\paren{\snr, h}. \label{Eqn: p(SNR, h) Lower Bound}
\end{eqnarray}
After scaling with $2^n$ and integrating \eqref{Eqn: p(SNR, h) Lower Bound} with respect to $f_{R, \Theta}\paren{r, \theta}$ over the above integration range, we have  
\begin{eqnarray}
p\paren{\snr} 
&\geq&  p_1\paren{\snr} + \frac12 p_2\paren{\snr} \nonumber \\
&=& L\paren{\snr}.  
\end{eqnarray}

Next, we establish that $U\paren{\snr} = p_1\paren{\snr} + 2p_2\paren{\snr}$ is an upper bound on $p\paren{\snr}$. To this end, we will show that $p_4\paren{\snr} \leq p_2\paren{\snr}$ for all $M \geq 4$. For $M=4$, this is trivial since $p_4\paren{\snr, h} = 0 \leq p_2\paren{\snr, h}$. For $M>4$, we define
\begin{eqnarray}
p_5\paren{\snr, h} = \frac{1}{\sqrt{\pi}}\int_{-\infty}^\infty \mathcal{Q}\paren{\sqrt{2\snr}r\sec\paren{\frac{2\pi}{M}}\sin\paren{\frac{2\pi}{M}-\theta} + \sqrt{2}w\tan\paren{\frac{2\pi}{M}}} \e{-w^2} dw. \nonumber 
\end{eqnarray}
We also define $p_5\paren{\snr}$ to be the integral of $2^n p_5\paren{\snr, h}$ with respect to $f_{R, \Theta}\paren{r, \theta}$ for $\theta \in \parenro{\frac{\pi}{M} - \frac{\pi}{2^n}, \frac{\pi}{M} + \frac{\pi}{2^n}}$ and $r \geq 0$.   

We observe that $p_4\paren{\snr} \leq p_5\paren{\snr}$ since the integrands are always positive and the integral with respect to $w$ is over the whole real line for $p_5\paren{\snr, h}$. Thus, it will be enough to show $p_2\paren{\snr} = p_5\paren{\snr}$ to conclude the proof. For $\snr = 0$, this can be verified by using the identity $\mathcal{Q}(x) = 1-\mathcal{Q}(-x)$. To prove the equality for all $\snr \geq 0$, we define the function $f\paren{\snr} = p_2\paren{\snr} - p_5\paren{\snr}$. It is enough to show that the derivative of $f\paren{\snr}$, which we represent by $f^\prime\paren{\snr}$, is equal to zero everywhere in order to show $p_2\paren{\snr} = p_5\paren{\snr}$. This is because if $f^\prime\paren{\snr}$ is equal to zero for all $\snr \geq 0$, then $f\paren{\snr}$ must be a constant function. Since $f\paren{0} = 0$, we have $f\paren{\snr} = p_2\paren{\snr} - p_5\paren{\snr} = 0$ for all $\snr \geq 0$. We devote the rest of the proof to showing that $f^\prime\paren{\snr} = 0$ for all $\snr \geq 0$.            

Using the definition of the $\mathcal{Q}$-function, the derivative of $p_2\paren{\snr}$ with respect to $\snr$, which we represent by $p_2^\prime\paren{\snr}$, is given by
\begin{eqnarray}
p_2^\prime\paren{\snr} &=& \frac{2^{n-1}}{\pi} \int_{\frac{\pi}{M} - \frac{\pi}{2^{n}}}^{\frac{\pi}{M} + \frac{\pi}{2^{n}}} \int_{0}^{\infty} \frac{d p_2\paren{\snr, r\e{\jmath \theta}}}{d\snr}  f_R(r) dr d\theta \nonumber \\
&=& \frac{2^{n-1}}{\pi} \int_{\frac{\pi}{M} - \frac{\pi}{2^{n}}}^{\frac{\pi}{M} + \frac{\pi}{2^{n}}} \int_{0}^{\infty} \frac{-r\sin \theta}{2 \sqrt{\pi \snr}} \e{-\snr r^2 \sin^2\theta} f_R(r) dr d\theta. \nonumber
\end{eqnarray}

Similarly, $p_5^\prime\paren{\snr}$ can be written as \begin{eqnarray} 
p_5^\prime\paren{\snr} &=& \frac{2^{n-1}}{\pi} \int_{\frac{\pi}{M} - \frac{\pi}{2^{n}}}^{\frac{\pi}{M} + \frac{\pi}{2^{n}}} \int_{0}^{\infty} \frac{d p_5\paren{\snr, r\e{\jmath \theta}}}{d\snr}  f_R(r) dr d\theta \nonumber \\
&=& \frac{2^{n-1}}{\pi \sqrt{\pi}} \int_{\frac{\pi}{M} - \frac{\pi}{2^{n}}}^{\frac{\pi}{M} + \frac{\pi}{2^{n}}} \int_{0}^{\infty} \frac{-A\paren{r, \theta}}{2\sqrt{\pi \snr}} I\paren{r, \theta} f_R(r)dr d\theta, \label{Eqn: p5(SNR) Derivative 1}
\end{eqnarray}
where $A\paren{r, \theta} = r \sec\paren{\frac{2\pi}{M}}\sin\paren{\frac{2\pi}{M} - \theta}$, $I\paren{r, \theta} = \int_{-\infty}^\infty \e{-\paren{w^2 + \paren{A\paren{r, \theta}\sqrt{\snr}+Bw}^2}}dw$ and $B= \tan\paren{\frac{2\pi}{M}}$. After completing the term in the exponent in $I\paren{r, \theta}$ to square and using the affinity of the resulting expression to a Gaussian pdf, $I\paren{r, \theta}$ can be shown to be equal to 
\begin{eqnarray}
I\paren{r, \theta} = \sqrt{\pi}\cos\paren{\frac{2\pi} {M}} \e{-\snr r^2 \sin^2\paren{\frac{2\pi}{M}-\theta}}. \label{Eqn: I(r, theta)}
\end{eqnarray}
Using \eqref{Eqn: I(r, theta)} in \eqref{Eqn: p5(SNR) Derivative 1}, we have 
\begin{eqnarray}
p_5^\prime\paren{\snr} = \frac{2^{n-1}}{\pi} \int_{\frac{\pi}{M} - \frac{\pi}{2^{n}}}^{\frac{\pi}{M} + \frac{\pi}{2^{n}}} \int_{0}^{\infty} \frac{-r\sin\paren{\frac{2\pi}{M}-\theta}}{2 \sqrt{\pi \snr}} \e{-\snr r^2 \sin^2\paren{\frac{2\pi}{M}-\theta}} f_R(r) dr d\theta. \label{Eqn: p5(SNR) Derivative 2}
\end{eqnarray}
Change of variables $u = \frac{2\pi}{M} - \theta$ in \eqref{Eqn: p5(SNR) Derivative 2} shows that $p_2^\prime\paren{\snr} = p_5^\prime\paren{\snr}$, and hence $f^\prime\paren{\snr} = 0$ as desired.   

\bibliographystyle{IEEEtran}
\bibliography{globecom_ref}

\begin{thebibliography}{10}
\providecommand{\url}[1]{#1}
\csname url@samestyle\endcsname
\providecommand{\newblock}{\relax}
\providecommand{\bibinfo}[2]{#2}
\providecommand{\BIBentrySTDinterwordspacing}{\spaceskip=0pt\relax}
\providecommand{\BIBentryALTinterwordstretchfactor}{4}
\providecommand{\BIBentryALTinterwordspacing}{\spaceskip=\fontdimen2\font plus
\BIBentryALTinterwordstretchfactor\fontdimen3\font minus
  \fontdimen4\font\relax}
\providecommand{\BIBforeignlanguage}[2]{{%
\expandafter\ifx\csname l@#1\endcsname\relax
\typeout{** WARNING: IEEEtran.bst: No hyphenation pattern has been}%
\typeout{** loaded for the language `#1'. Using the pattern for}%
\typeout{** the default language instead.}%
\else
\language=\csname l@#1\endcsname
\fi
#2}}
\providecommand{\BIBdecl}{\relax}
\BIBdecl

\bibitem{Bai15}
Q.~Bai and J.~A. Nossek, ``Energy efficiency maximization for 5{G}
  multi-antenna receivers,'' \emph{Trans. Emerging Telecommunications
  Technologies}, vol.~26, pp. 3--14, 2015.

\bibitem{paper_40}
S.~Rangan, T.~S. Rappaport, and E.~Erkip, ``Millimeter-wave cellular wireless
  networks: Potentials and challenges,'' \emph{Proc. IEEE}, vol. 102, no.~3,
  pp. 366--385, Mar. 2014.

\bibitem{paper_46}
\BIBentryALTinterwordspacing
B.~Murmann, ``{ADC} performance survey 1997-2017.'' [Online]. Available:
  \url{http://web.stanford.edu/~murmann/adcsurvey.htm}
\BIBentrySTDinterwordspacing

\bibitem{paper_55}
J.~Zhang, L.~Dai, X.~Li, Y.~Liu, and L.~Hanzo, ``On low-resolution {ADC}s in
  practical 5{G} millimeter-wave massive {MIMO} systems,'' \emph{IEEE Commun.
  Mag.}, vol.~56, no.~7, pp. 205--211, Jul. 2018.

\bibitem{paper_22}
M.~T. Ivrlac and J.~A. Nossekh, ``On {MIMO} channel estimation with single-bit
  signal-quantization,'' in \emph{Proc. Int. ITG/IEEE Workshop on Smart
  Antennas (WSA)}, Vienna, Austria, Feb. 2007.

\bibitem{paper_23}
T.~M. Lok and V.~K.-W. Wei, ``Channel estimation with quantized observations,''
  in \emph{Proc. 1998 IEEE International Symposium on Information Theory}, Aug.
  1998, pp. 333--333.

\bibitem{Madow09}
J.~Singh, S.~Ponnuru, and U.~Madhow, ``Multi-gigabit communication: the adc
  bottleneck1,'' in \emph{2009 IEEE International Conference on
  Ultra-Wideband}, Vancouver, Canada, Sep. 2009, pp. 22--27.

\bibitem{Dabeer10}
O.~Dabeer and U.~Madhow, ``Channel estimation with low-precision
  analog-to-digital conversion,'' in \emph{2010 IEEE International Conference
  on Communications}, Cape Town, South Africa, May 2010, pp. 1--6.

\bibitem{paper_20}
J.~Mo, P.~Schniter, N.~G. Prelcic, and R.~W. Heath, ``Channel estimation in
  millimeter wave {MIMO} systems with one-bit quantization,'' in \emph{Proc.
  48th Asilomar Conference on Signals, Systems and Computers}, Pacific Grove,
  California, Nov. 2014, pp. 957--961.

\bibitem{IoT5g}
\BIBentryALTinterwordspacing
``Mobile {I}o{T} in the 5{G} {F}uture.'' [Online]. Available:
  \url{https://www.ericsson.com/assets/local/networks/documents/gsma-5g-mobile-iot.pdf}
\BIBentrySTDinterwordspacing

\bibitem{paper_3}
J.~Choi, D.~J. Love, D.~R. Brown, and M.~Boutin, ``Quantized distributed
  reception for {MIMO} wireless systems using spatial multiplexing,''
  \emph{IEEE Trans. Signal Process.}, vol.~63, no.~13, pp. 3537--3548, Jul.
  2015.

\bibitem{paper_8}
J.~Choi, J.~Mo, and R.~W. Heath, ``Near maximum-likelihood detector and channel
  estimator for uplink multiuser massive {MIMO} systems with one-bit {ADC}s,''
  \emph{IEEE Trans. Commun.}, vol.~64, no.~5, pp. 2005--2018, May 2016.

\bibitem{paper_25}
J.~Choi, D.~J. Love, and D.~R. Brown, ``Channel estimation techniques for
  quantized distributed reception in {MIMO} systems,'' in \emph{Proc. 48th
  Asilomar Conference on Signals, Systems and Computers}, Pacific Grove,
  California, Nov. 2014, pp. 1066--1070.

\bibitem{Lee17}
Y.~Li, C.~Tao, G.~Seco-Granados, A.~Mezghani, A.~L. Swindlehurst, and L.~Liu,
  ``Channel estimation and performance analysis of one-bit massive {MIMO}
  systems,'' \emph{IEEE Trans. Signal Process.}, vol.~65, no.~15, pp.
  4075--4089, Aug 2017.

\bibitem{paper_32}
A.~K. Saxena, I.~Fijalkow, and A.~L. Swindlehurst, ``On one-bit quantized {ZF}
  precoding for the multiuser massive {MIMO} downlink,'' in \emph{Proc. 2016
  IEEE Sensor Array and Multichannel Signal Processing Workshop (SAM)}, Rio de
  Janeiro, Brazil, Jul. 2016, pp. 1--5.

\bibitem{paper_34}
------, ``Analysis of one-bit quantized precoding for the multiuser massive
  {MIMO} downlink,'' \emph{IEEE Trans. Signal Process.}, vol.~65, no.~17, pp.
  4624--4634, Sep. 2017.

\bibitem{paper_33}
A.~Swindlehurst, A.~Saxena, A.~Mezghani, and I.~Fijalkow, ``Minimum
  probability-of-error perturbation precoding for the one-bit massive {MIMO}
  downlink,'' in \emph{Proc. 2017 IEEE International Conference on Acoustics,
  Speech and Signal Processing (ICASSP)}, New Orleans, USA, Mar. 2017, pp.
  6483--6487.

\bibitem{paper_26}
J.~Mo and R.~W. Heath, ``High {SNR} capacity of millimeter wave {MIMO} systems
  with one-bit quantization,'' in \emph{Proc. 2014 Information Theory and
  Applications Workshop (ITA)}, San Diego, California, USA, Feb. 2014, pp.
  1--5.

\bibitem{paper_29}
------, ``Capacity analysis of one-bit quantized {MIMO} systems with
  transmitter channel state information,'' \emph{IEEE Trans. Signal Process.},
  vol.~63, no.~20, pp. 5498--5512, Oct. 2015.

\bibitem{paper_27}
A.~Mezghani and J.~A. Nossek, ``Analysis of 1-bit output noncoherent fading
  channels in the low {SNR} regime,'' in \emph{Proc. 2009 IEEE International
  Symposium on Information Theory}, Seoul, Korea, Jun. 2009, pp. 1080--1084.

\bibitem{paper_28}
------, ``On ultra-wideband {MIMO} systems with 1-bit quantized outputs:
  Performance analysis and input optimization,'' in \emph{Proc. 2007 IEEE
  International Symposium on Information Theory}, Nice, France, Jun. 2007, pp.
  1286--1289.

\bibitem{paper_30}
Y.~Li, C.~Tao, A.~L. Swindlehurst, A.~Mezghani, and L.~Liu, ``Downlink
  achievable rate analysis in massive {MIMO} systems with one-bit {DAC}s,''
  \emph{IEEE Commun. Lett.}, vol.~21, no.~7, pp. 1669--1672, Jul. 2017.

\bibitem{paper_31}
J.~Singh and U.~Madhow, ``Phase-quantized block noncoherent communication,''
  \emph{IEEE Trans. Commun.}, vol.~61, no.~7, pp. 2828--2839, Jul. 2013.

\bibitem{paper_36}
N.~Liang and W.~Zhang, ``Mixed-{ADC} massive {MIMO},'' \emph{IEEE J. Sel. Areas
  Commun.}, vol.~34, no.~4, pp. 983--997, April 2016.

\bibitem{Mezghani_12}
A.~Mezghani, M.~S. Khoufi, and J.~A. Nossek, ``A modified {MMSE} receiver for
  quantized {MIMO} systems,'' in \emph{Proc. Int. ITG/IEEE Workshop on Smart
  Antennas (WSA)}, Vienna, Austria, 2007.

\bibitem{paper_13}
A.~Mezghani, M.~Rouatbi, and J.~A. Nossek, ``An iterative receiver for
  quantized {MIMO} systems,'' in \emph{Proc. 2012 16th IEEE Mediterranean
  Electrotechnical Conference}, Yasmine Hammamet, Tunisia, Mar. 2012, pp.
  1049--1052.

\bibitem{Krone10}
S.~Krone and G.~Fettweis, ``Fading channels with 1-bit output quantization:
  Optimal modulation, ergodic capacity and outage probability,'' in \emph{2010
  IEEE Information Theory Workshop}, Dublin, Ireland, Aug 2010, pp. 1--5.

\bibitem{book_5}
R.~G. Gallager, \emph{Principles of Digital Communication}.\hskip 1em plus
  0.5em minus 0.4em\relax New York, NY, USA: Cambridge University Press, 2008.

\bibitem{Remmert91}
R.~Remmert, \emph{Theory of Complex Functions}.\hskip 1em plus 0.5em minus
  0.4em\relax New York: Springer-Verlag, 1991.

\bibitem{Picinbono94}
B.~Picinbono, ``On circularity,'' \emph{IEEE Trans. Signal Process.}, vol.~42,
  no.~12, pp. 3473--3482, Dec. 1994.

\bibitem{Koivunen12}
E.~Ollila, D.~E. Tyler, V.~Koivunen, and H.~V. Poor, ``Complex elliptically
  symmetric distributions: {Survey}, new results and applications,'' \emph{IEEE
  Trans. Signal Process.}, vol.~60, no.~11, pp. 5597--5625, Nov. 2012.

\bibitem{Mezghani2010}
A.~Mezghani, F.~Antreich, and J.~A. Nossek, ``Multiple parameter estimation
  with quantized channel output,'' in \emph{Proc. 2010 International ITG
  Workshop on Smart Antennas (WSA)}, Feb 2010, pp. 143--150.

\bibitem{Mo2018}
J.~Mo, P.~Schniter, and R.~W. Heath, ``Channel estimation in broadband
  millimeter wave {MIMO} systems with few-bit {ADC}s,'' \emph{IEEE Trans.
  Signal Process.}, vol.~66, no.~5, pp. 1141--1154, March 2018.

\bibitem{Wen16}
C.~Wen, C.~Wang, S.~Jin, K.~Wong, and P.~Ting, ``Bayes-optimal joint
  channel-and-data estimation for massive {MIMO} with low-precision {ADC}s,''
  \emph{IEEE Trans. Signal Process.}, vol.~64, no.~10, pp. 2541--2556, May
  2016.

\bibitem{Cover91}
T.~M. Cover and J.~A. Thomas, \emph{Elements of Information Theory}.\hskip 1em
  plus 0.5em minus 0.4em\relax New York, NY, USA: John Wiley \& Sons, 1991.

\bibitem{Massey93}
F.~D. Neeser and J.~L. Massey, ``Proper complex random processes with
  applications to information theory,'' \emph{IEEE Trans. Inf. Theory},
  vol.~39, no.~4, pp. 1293--1302, Jul. 1993.

\bibitem{Nakagami60}
M.~Nakagami, ``The $m$-distribution - {A} general formula of intensity
  distribution of rapid fading,'' \emph{Statistical Methods in Radio Wave
  Propagation - Pergamon Press}, pp. 7--36, 1960.

\bibitem{Stuber:2001}
G.~L. St\"{u}ber, \emph{Principles of Mobile Communication (2Nd Ed.)},
  2nd~ed.\hskip 1em plus 0.5em minus 0.4em\relax Norwell, MA, USA: Kluwer
  Academic Publishers, 2001.

\bibitem{book_1}
I.~S. Gradshteyn and I.~M. Ryzhik, \emph{Table of integrals, series, and
  products}, 7th~ed.\hskip 1em plus 0.5em minus 0.4em\relax Elsevier/Academic
  Press, Amsterdam, 2007.

\bibitem{paper_49}
N.~C. Beaulieu and C.~Cheng, ``Efficient nakagami-m fading channel
  simulation,'' \emph{IEEE Trans. Veh. Technol.}, vol.~54, no.~2, pp. 413--424,
  Mar. 2005.

\bibitem{book_7}
W.~Rudin, \emph{Real and Complex Analysis}, 3rd~ed.\hskip 1em plus 0.5em minus
  0.4em\relax New York: McGraw-Hill, 1987.

\bibitem{paper_12}
R.~H. Walden, ``Analog-to-digital converter survey and analysis,'' \emph{IEEE
  J. Sel. Areas Commun.}, vol.~17, no.~4, pp. 539--550, Apr. 1999.

\bibitem{Lucky62}
R.~W. Lucky and J.~C. Hancock, ``On the optimum performance of $n$-ary systems
  having two-degrees of freedom,'' \emph{IRE Trans. Commun. Syst.}, vol.~10,
  no.~2, pp. 185--192, Jun. 1962.

\bibitem{Tsitsiklis08}
D.~P. Bertsekas and J.~N. Tsitsiklis, \emph{Introduction to Probability},
  2nd~ed.\hskip 1em plus 0.5em minus 0.4em\relax Nashua: Athena Scientific,
  2008.

\bibitem{book_8}
K.~T. Fang, S.~Kotz, and K.~W. Ng, \emph{Symmetric Multivariate and Related
  Distributions}, 1st~ed.\hskip 1em plus 0.5em minus 0.4em\relax New York: CRC
  Press, 1990.

\end{thebibliography}

\clearpage
\end{document}